\documentclass[letterpaper, paper,11pt]{AAS}	

\usepackage{bm}
\usepackage{amsmath}
\usepackage{amsfonts}
\usepackage{amsthm}
\usepackage{svg}
\usepackage{subfigure}
\usepackage[ruled]{algorithm2e}
\usepackage[colorlinks=true, pdfstartview=FitV, linkcolor=black, citecolor= black, urlcolor= black]{hyperref}
\usepackage{overcite}
\usepackage{footnpag}			      	
\usepackage{xcolor}

\newtheorem{problem}{Problem}

\PaperNumber{25-549}

\begin{document}

\title{Robust Sampling-Based Covariance Steering for Aerocapture Guidance}

\author{Alex Rose\thanks{PhD Student, Department of Aeronautics and Astronautics, Massachusetts Institute of Technology, Cambridge, MA, 02139, USA; Draper Scholar, The Charles Stark Draper Laboratory, Inc., Cambridge, MA, 02139, USA.},  
Christopher Jewison \thanks{Aerospace Engineer, The Charles Stark Draper Laboratory, Inc., Cambridge, MA, 02139, USA.},
\ and Jonathan P. How\thanks{Richard C. Maclaurin Professor in Aeronautics and Astronautics, Massachusetts Institute of Technology, Cambridge, MA, 02139, USA.}
}

\maketitle{}

\begin{abstract}
Aerocapture is a maneuver where a spacecraft dives through the atmosphere of a planet or moon to reduce its velocity and prepare for orbital insertion. Aerocapture allows for higher cruise velocities and reduces fuel consumption, decreasing transit time and increasing payload mass. However, uncertainties in the atmospheric entry state and atmospheric density increase the risk of aerocapture. Dynamic nonlinearities and nonlinearities caused by the state-dependence of the atmospheric density pose additional challenges. This work develops a robust sampling-based covariance steering algorithm designed for aerocapture guidance. Our proposed algorithm leverages sampled nonlinear system trajectories to improve evaluation of the $\Delta V$ required for aerocapture and address nonlinearities caused by the aerocapture dynamics and atmospheric disturbances. We perform Monte Carlo simulations with dispersed entry and atmospheric conditions on aerocapture scenarios at Mars and Uranus and demonstrate a 5-15\% reduction in the 99th-percentile, 99.7th-percentile, and worst-case $\Delta V$ required for aerocapture when compared against a state-of-the-art covariance steering algorithm.
\end{abstract}

\section{Introduction}
The 2023-2023 Planetary Science Decadal Survey \cite{national2022origins} suggests that Mars Sample Return (MSR) should be prioritized above all other robotic exploration missions in the next decade. The Perseverance rover has already collected high-quality rock samples from Jezero Crater which are key to understanding Mars' geologic history \cite{national2022origins}. Returning samples from Mars is technically and operationally challenging and requires delivering an extremely heavy payload to Mars. The 2023-2023 Planetary Science Decadal Survey also proposes a high-priority flagship mission to Uranus. The proposed Uranus Orbiter and Probe mission presents a valuable opportunity to investigate the formation of the ice giants and learn about the atmosphere, axial tilt, and magnetic field  of Uranus. However, Uranus' distance from Earth means that a mission relying on traditional fully propulsive orbital insertion would require a 13-15 year cruise time \cite{national2022origins, dutta2020aerocapture}. 

\begin{figure}[t]
	\centering\includegraphics[width=5in]{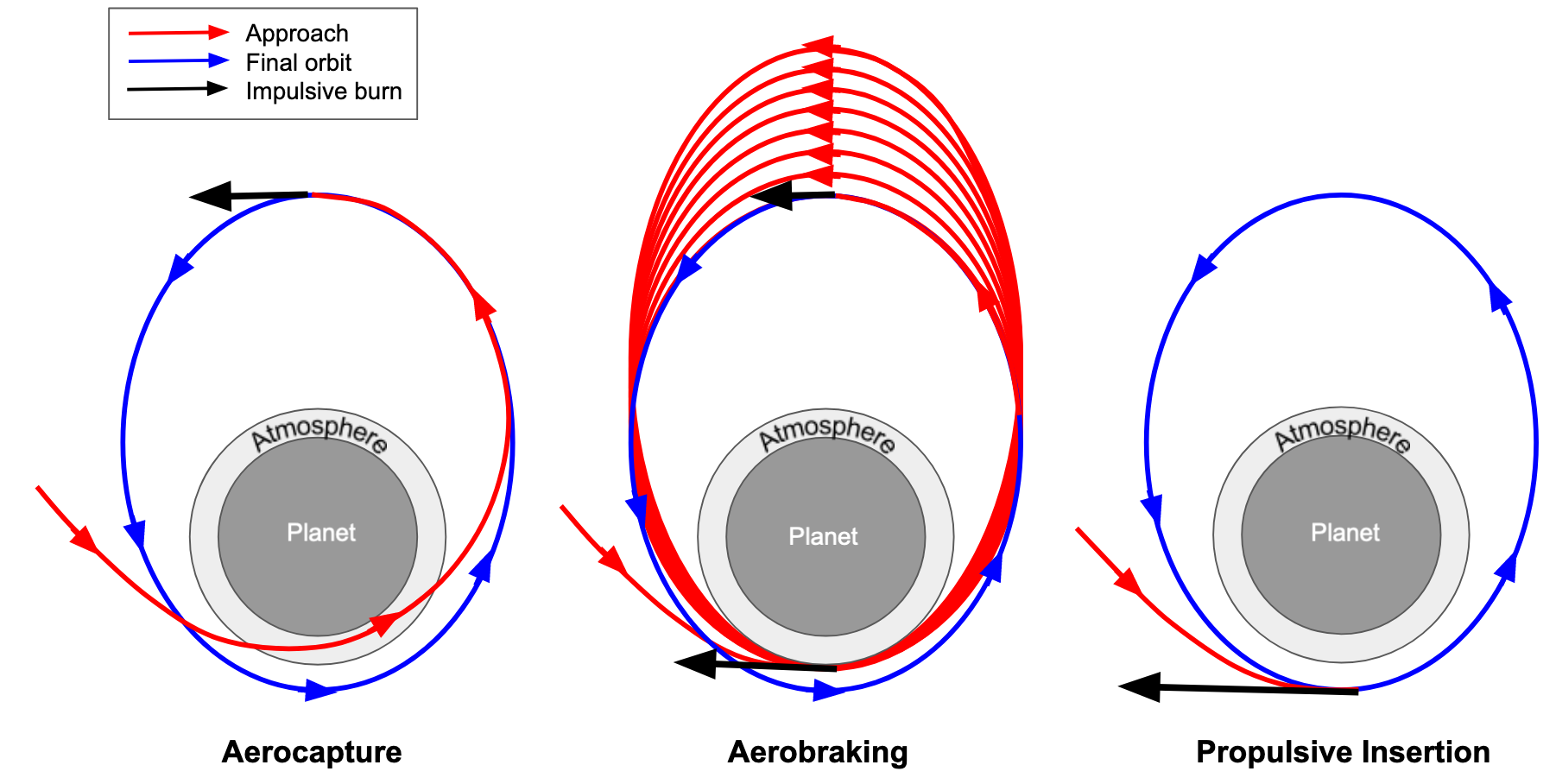}
	\caption{Aerocapture (left) uses atmospheric drag to reduce spacecraft velocity, while alternatives require a large impulsive burn to slow down enough to enter into an elliptical orbit.}
	\label{fig:aerocapture_alternatives}
\end{figure}

Aerocapture, a maneuver where a spacecraft dives through the atmosphere of a planet or moon to reduce its velocity and prepare for orbital insertion, is key to enabling long-duration missions with heavy payloads. Aerocapture can provide a larger velocity reduction than can be achieved with fully propulsive orbital insertion, enabling a higher cruise velocity and reducing transit time to Uranus by 2-5 years \cite{dutta2020aerocapture}. Prior studies on aerocapture \cite{wright2006mars} also suggest that aerocapture at Mars reduces fuel consumption enough to decrease the required launch mass for a sample return mission by 3-4 times, enabling heavier payloads to reach Mars with existing launch vehicles. The two major alternatives to aerocapture are fully propulsive orbital insertion and aerobraking, where a spacecraft uses an impulsive burn to enter a highly eccentric elliptical orbit and repeatedly passes through the upper atmosphere to lower its apoapsis. Both aerobraking and fully propulsive orbital insertion require a large impulsive burn to enter an elliptical orbit around a planet (see Figure \ref{fig:aerocapture_alternatives}). Aerocapture, in contrast, uses atmospheric drag to slow down a hyperbolic approach trajectory and capture into an elliptical orbit, saving fuel and increasing payload mass by 40\% when compared to fully propulsive orbital insertion \cite{dutta2020aerocapture}.

Significant atmospheric uncertainty presents a major challenge for aerocapture \cite{spilker2019qualitative}. UranusGRAM, the state-of-the-art atmospheric model for Uranus, relies heavily on Voyager 2 observations of Uranus from the 1980's and includes significant uncertainties in atmospheric density perturbations \cite{justh2024uranus}. Mars' atmosphere is better understood, but predicting atmospheric density and heat flux during hypersonic flight remains challenging \cite{albert2025dimensionality}. State-of-the-art deterministic aerocapture guidance algorithms \cite{lu2015optimal, rataczak2025predictor, sonandres2025aerocapture} adopt a predictor-corrector architecture, where the prediction phase integrates the vehicle dynamics forward through the atmosphere, and the correction phase computes an error metric and updates the guidance trajectory to minimize the error metric. As such, predictor-corrector guidance algorithms are highly sensitive to the atmospheric profile used in the prediction phase, and can fail when the true atmospheric density does not match the atmospheric density profile used by guidance. Even when atmospheric density is estimated online, ``high-to-low'' density scenarios, where atmospheric drag starts higher than expected and then suddenly drops, cause predictor-correctors to overestimate the control authority available and thus can cause large orbit misses \cite{matz2024analysis}.

Robust guidance algorithms seek to reduce the sensitivity of aerocapture trajectories to the atmospheric profile used by guidance to propagate the vehicle dynamics. Desensitized aerocapture guidance \cite{chadalavada2025desensitized} augments the guidance objective to reduce sensitivity to atmospheric error, transforming the aerocapture guidance problem into a nonlinear problem requiring a pseudospectral solver. $\pi$PAG trains a Gaussian Mixture Variational Autoencoder (GMVAE) to estimate the probabilities of escape and successful aerocapture, and corrects guidance trajectories when there is a high probability of escape predicted by the GMVAE \cite{calkinsrisk}. However, these algorithms do not explicitly propagate or account for state uncertainty resulting from atmospheric uncertainty.

Chance-constrained covariance steering explicitly models atmospheric uncertainty as a Gaussian random field and optimizes over the probability distribution of aerocapture exit states \cite{ridderhof2022chance}. Empirical atmospheric models can be modeled as Gaussian random fields using Karhunen-Lo\`eve expansion, or by simply adopting the unbiased sample mean and covariance \cite{albert2023onboard}. Although prior work on covariance steering for aerocapture\cite{ridderhof2022chance} considers atmospheric uncertainty, this work does not account for nonlinearities caused by the state dependence of the atmospheric uncertainty or by the nonlinear aerocapture dynamics. Our prior work \cite{rose2024revise} accounts for nonlinearities caused by state-dependent uncertainties by modeling the state distribution as a mixture of Gaussians and develops a robust sampling-based method for covariance steering for systems affected by a state-dependent Gaussian random field, but only considers linear dynamics with a covariance-minimizing objective. Recent advances in uncertainty quantification for aerocapture also typically model the state distribution with a Gaussian mixture model, and explore nonlinear covariance propagation techniques such as using dynamics-informed directional state transition tensors \cite{calkinsdynamics}.

In this work, we develop a new robust sampling-based covariance steering algorithm that minimizes a nonlinear objective for a nonlinear system affected by a state-dependent environmental disturbance. Our proposed algorithm samples a collection of initial states, rolls out nonlinear system trajectories, and minimizes the worst-case value of the objective function over the collection of trajectories. Our sampling-based approach improves robustness to initial state error by improving modeling accuracy for dynamic nonlinearities, nonlinearities caused by state-dependent uncertainties, and nonlinearities in the objective function. We demonstrate our robust sampling-based covariance steering algorithm on three aerocapture scenarios: an easy scenario at Mars with a small initial state dispersion (modified from Ridderhof \& Tsiotras \cite{ridderhof2022chance}), a more challenging scenario at Mars with a large initial velocity dispersion, and a realistic scenario at Uranus with a dispersed initial state and a highly elliptical target orbit (modified from Matz et al. \cite{matz2024analysis}). Across all scenarios, our robust sampling-based algorithm reduces the 99th percentile $\Delta V$, 99.7th-percentile $\Delta V$, and worst-case $\Delta V$ required for aerocapture by 5-15\% when compared to a state-of-the-art covariance steering algorithm \cite{ridderhof2022chance}.
\section{Problem Statement}
\subsection{Aerocapture Dynamics}
As in Lu et al.\cite{lu2015optimal} and Ridderhof \& Tsiotras \cite{ridderhof2022chance}, we focus on longitudinal aerocapture guidance in this paper, as lateral aerocapture guidance can typically be achieved with occasional bank angle reversals. Consider the longitudinal aerocapture dynamics
\begin{align}
\dot{r} &= v \sin \gamma, \notag \\
\dot{v} &= -\frac{\rho(r) v^2 }{2 B_c} - \frac{\mu \sin \gamma}{r^2}, \label{eq: aerocapture_dynamics} \\
\dot{\gamma} &= \frac{\rho(r) v (L/D)}{2 B_c} \cos \sigma - \left(\frac{\mu}{r^2} - \frac{v^2}{r}\right)\frac{\cos \gamma}{v}, \notag
\end{align}
where $r$ represents the orbital radius, $v$ represents the orbital velocity, $\gamma$ represents the flight-path angle, $\sigma$ represents the vehicle bank angle, $\mu$ is the gravitational constant, $\rho(r)$ is the atmospheric density, $B_c$ is the ballistic coefficient of the vehicle, and $L/D$ is the lift-to-drag ratio of the vehicle.

Upon atmospheric exit, successful aerocapture requires an impulsive burn to raise the orbit periapsis out of the atmosphere to a target periapsis $r_{p, \textrm{goal}}$. This burn requires $\Delta V$ equal to:
\begin{equation}
\Delta V_1 = \sqrt{2 \mu}\left(\sqrt{\left(\frac{1}{r_{a, \textrm{exit}}} - \frac{1}{r_{a, \textrm{exit}}+ r_{p, \textrm{goal}}} \right)} -  \sqrt{\left(\frac{1}{r_{a, \textrm{exit}}} - \frac{1}{2a} \right)}\right),
\end{equation}
where the vehicle exit apoapsis $r_{a, \textrm{exit}}$ is given by 
\begin{equation}
r_{a, \textrm{exit}} = a\left(1 + \sqrt{1 - \frac{v_{\textrm{exit}}^2 r_{\textrm{exit}}^2 \cos^2(\gamma_{\textrm{exit}})}{\mu a}}\right)
\end{equation}
and the semimajor axis $a$ at exit is given by 
\begin{equation}
a = \frac{\mu}{2\mu/r_\textrm{exit} - v_\textrm{exit}^2}.
\end{equation}
If the vehicle exit apoapsis $r_{a, \textrm{exit}}$ does not exactly match the target orbit apoapsis $r_{a, \textrm{goal}}$, a second burn is required to reach the target apoapsis. The $\Delta V$ required for the apoapsis cleanup burn is equal to
\begin{equation}
\Delta V_2 = \sqrt{2 \mu}\left|\left(\sqrt{\left(\frac{1}{r_{p, \textrm{goal}}} - \frac{1}{r_{a, \textrm{goal}} + r_{p, \textrm{goal}}} \right)} - \sqrt{\left(\frac{1}{r_{p, \textrm{goal}}} - \frac{1}{r_{a, \textrm{exit}} + r_{p, \textrm{goal}}} \right)}\right)\right|,
\end{equation}
and may be zero in the case of perfect apoapsis targeting. The total $\Delta V$ required for successful aerocapture is given by $\Delta V_1 + \Delta V_2$, and is a function of the atmospheric exit state and the periapsis and apoapsis of the target orbit.
\subsection{Stochastic Problem Formulation}
With $\mathbf{x} = [r, v, \gamma]^T$, $\mathbf{u} = [\cos \sigma]^T$, $\Psi(\phi(\mathbf{x})) = [\rho(r)]^T$, the continuous-time aerocapture dynamics given in Eq. \ref{eq: aerocapture_dynamics} can be expressed in vector form as $\dot{\mathbf{x}}(t) = f(\mathbf{x}(t), \mathbf{u}(t), \Psi(\phi(\mathbf{x}(t))))$. 
We presume a Gaussian initial state dispersion, such that the initial state $\mathbf{x}_0 \sim \mathcal{N}(\mu_0, \Sigma_0)$. We also model the altitude-dependent atmospheric density $\rho(r)$ by a Gaussian random field $\Psi(\phi(x))$, with a mean function $\overline{\Psi}(\phi(\mathbf{x}))$ and a covariance function $\Sigma_{\Psi}(\phi(\mathbf{x}_i), \phi(\mathbf{x}_j))$. Given our probabilistic problem formulation, we can reformulate bank angle constraints as probabilistic chance constraints of the form $\mathbb{P}(\mathbf{u}_k \in \mathcal{U}) \geq 1 - \epsilon_u$, where $\mathcal{U}$ is the set of allowable bank angles, and $\epsilon_u$ is the probability of constraint violation. 

We seek to minimize the $X$th-percentile $\Delta V$ required for successful aerocapture (generally with $X = 99$ or $X = 99.7$). As such, we define the following stochastic optimization problem, where $\mathbf{x}_f$ indicates the state at atmospheric exit:
\begin{problem} \label{prob: aerocapture_delta_V}
\textit{Minimize the $X$th-percentile $\Delta V$ required for successful aerocapture, subject to probabilistic bank angle constraints, with a Gaussian initial state distribution and with the atmospheric density modeled by a state-dependent Gaussian random field.}
\begin{equation}
\inf J = \{\gamma \in \mathbb{R}: \mathbb{P}(\Delta V(\mathbf{x}_f) > \gamma) \leq X/100) \}
\end{equation}
such that:
\begin{equation}
\begin{split}
&\dot{\mathbf{x}} = f(\mathbf{x}, \mathbf{u}, \Psi(\phi(\mathbf{x}))) \\
&\mathbf{x}_0 \sim \mathcal{N}(\mu_0, \Sigma_0),\ \Psi \sim \mathcal{N}(\overline{\Psi}, \Sigma_\Psi),\\
&\mathbb{P}(\mathbf{u}_k \in \mathcal{U}) \geq 1 - \epsilon_u.
\end{split}
\end{equation}
\end{problem}
\section{Approach}
The stochastic optimization problem in Problem \ref{prob: aerocapture_delta_V} is nonlinear and is challenging to solve directly. As in Ridderhof \& Tsiotras\cite{ridderhof2022chance}, we adopt a successive convexification approach, reformulating Problem \ref{prob: aerocapture_delta_V} as a convex covariance steering problem that can be solved efficiently by existing commercial solvers.

\subsection{Iterative Nonlinear Covariance Steering in a Gaussian Random Field}
We use the notation that a vector $\mathbf{V}$ represents a column of stacked $\mathbf{v}_k$ for all $k=0, \ldots, N$, $\overline{\mathbf{V}} = \mathbb{E}[\mathbf{V}]$, and that $\widetilde{\mathbf{V}} = \mathbf{V}- \overline{\mathbf{V}}$. We also use the notation that for a continuous time-varying trajectory $\hat{\mathbf{v}}(t)$, $\hat{\mathbf{v}}_k = \hat{\mathbf{v}}(t_k)$.

Suppose we have a nominal control input $\hat{\mathbf{u}}$ on the time interval $[t_0, t_f]$. Such an input can be generated by a deterministic optimal control algorithm. We propagate the nominal state trajectory according to the system dynamics
\begin{equation}\label{eq: nominal_trajectory_propagation}
\dot{\hat{\mathbf{x}}} = f(\hat{\mathbf{x}}, \hat{\mathbf{u}}, \mathbb{E}[\Psi(\phi(\hat{\mathbf{x}}))])
\end{equation}
and evaluate the nominal disturbance trajectory $\hat{\Psi}(t) = \Psi(\phi(\hat{\mathbf{x}(t)}))$.
Next, we discretize and linearize the dynamics about the nominal state, control, and disturbance trajectories given by $\hat{\mathbf{x}}(t), \hat{\mathbf{u}}(t)$, and $\hat{\Psi}(t)$. Then, given a set of discrete timesteps $[t_0, \ldots, t_k, \ldots, t_N]$, with $t_N = t_f$, we have that for all $k < N$:
\begin{equation}\label{eq: discrete_time_dynamics}
\mathbf{x}_{k+1} = A_k\mathbf{x}_k + B_k\mathbf{u}_k + \mathbf{c}_k + G_k\mathbf{w}_k,
\end{equation}
where 
\begin{equation}
A_k =  \frac{\partial}{\partial \hat{\mathbf{x}}_k}\int_{t_k}^{t_{k+1}} f(\hat{\mathbf{x}}(t), \hat{\mathbf{u}}(t), \hat{\Psi}(t))\ dt,
\end{equation}
\begin{equation}
B_k  =  \frac{\partial}{\partial \hat{\mathbf{u}}_k}\int_{t_k}^{t_{k+1}} f(\hat{\mathbf{x}}(t), \hat{\mathbf{u}}(t), \hat{\Psi}(t))\ dt,
\end{equation}
\begin{equation}
\mathbf{c}_k = \hat{\mathbf{x}}_{k+1} - A_k\hat{\mathbf{x}}_k - B_k\hat{\mathbf{u}}_k - G_k \hat{\Psi}_k,
\end{equation}
\begin{equation}
G_k = \frac{\partial}{\partial \hat{\Psi}}\int_{t_k}^{t_{k+1}} f(\hat{\mathbf{x}}(t), \hat{\mathbf{u}}(t), \hat{\Psi}(t))\ dt,
\end{equation}
\begin{equation}\label{eq: discretize_w}
\mathbf{w}_k = \Psi(\phi(\mathbf{x}_k))
\end{equation}
Because $\hat{\Psi}(t)$ depends on $\hat{\mathbf{x}}(t)$, $A_k$ will depend partially on $\hat{\Psi}(t)$ for each $k$. We approximate the state-dependent Gaussian random field disturbance $\Psi(\phi(\mathbf{x}(t)))$ as a time-varying disturbance $\mathbf{W} \sim \mathcal{N}(\overline{\mathbf{W}}, \Sigma_W)$ by discretizing it around the nominal state trajectory \cite{ridderhof2022chance, rose2024revise}, with $\overline{\mathbf{W}} = \mathbb{E}[\Psi(\phi(\hat{\mathbf{X}}))]$ and $\Sigma_W$ such that $\Sigma_{W_{i, j}} = \Sigma_\Psi(\phi(\hat{\mathbf{x}}_i, \hat{\mathbf{x}}_j)$ $\forall i, j$.

We express the discrete-time dynamics given in Eq. \ref{eq: discrete_time_dynamics} in block-matrix form as \cite{okamoto2018optimal}
\begin{equation}\label{eq: block_matrix_dynamics}
\begin{bmatrix} \mathbf{x}_0 \\ \mathbf{x}_1 \\ \mathbf{x}_2 \\ \vdots \end{bmatrix} = \begin{bmatrix}I \\ A_0 \\ A_1A_0 \\ \vdots \end{bmatrix}\mathbf{x}_0 + \begin{bmatrix}0 & 0 & \\ B_0 & 0 & \\ A_1B_0 & B_1 & \\ && \ddots \end{bmatrix}\begin{bmatrix}\mathbf{u}_0 \\ \mathbf{u}_1 \\ \vdots \end{bmatrix} + \begin{bmatrix}0 \\ \mathbf{c}_0 \\ A_1\mathbf{c}_1 \\ \vdots \end{bmatrix} + \begin{bmatrix}0 & 0 & \\ G_0 & 0 & \\ A_1G_0 & G_1 & \\ && \vdots \end{bmatrix}\begin{bmatrix}\mathbf{w}_0 \\ \mathbf{w}_1 \\ \vdots \end{bmatrix}.
\end{equation}
Using the notation that a vector $\mathbf{V}$ represents a column of stacked $\mathbf{v}_k$ for all $k=0, \ldots, N$, Eq. \ref{eq: block_matrix_dynamics} can be expressed by
\begin{equation}
\mathbf{X} = A\mathbf{x}_0 + B\mathbf{U} + \mathbf{C} + G\mathbf{W}.
\end{equation}
and so the mean state dynamics are given by
\begin{equation}\label{eq: mean_state_dynamics}
\overline{\mathbf{X}} = A\overline{\mathbf{x}}_0 + B\overline{\mathbf{U}} + \mathbf{C} + G\overline{\mathbf{W}}.
\end{equation}
We use a state history feedback control law \cite{ridderhof2022chance, rose2024revise}, such that
\begin{equation}
\mathbf{u}_k = \sum_{i=0}^k K_{k, i}(\mathbf{x}_k - \overline{\mathbf{x}}_k) + \overline{\mathbf{u}}_k,
\end{equation}
with $\overline{\mathbf{u}}_k$ the nominal control at time $k$. This control law can be expressed in block-matrix form as
\begin{equation}
\mathbf{U} = K(\mathbf{X} - \overline{\mathbf{X}}) + \overline{\mathbf{U}}.
\end{equation}
It follows that
\begin{equation}
\widetilde{\mathbf{X}} = \mathbf{X} - \overline{\mathbf{X}} = A(\mathbf{x}_0 - \overline{\mathbf{x}_0}) + BK\widetilde{\mathbf{X}} + G\widetilde{\mathbf{W}},
\end{equation}
and that the state covariance evolves according to
\begin{equation}\label{eq: covariance_dynamics}
\text{Cov}(\mathbf{X}) = \mathbb{E}[\widetilde{\mathbf{X}}\widetilde{\mathbf{X}}^T] = (I - BK)^{-1}(A\Sigma_0A^T + G\Sigma_W G^T){(I - BK)^{-1}}^T.
\end{equation}
We see also that
\begin{equation}
\text{Cov}(\mathbf{U}) = K\mathbb{E}[\widetilde{\mathbf{X}}\widetilde{\mathbf{X}}^T]K^T = K(I - BK)^{-1}(A\Sigma_0A^T + G\Sigma_W G^T){(I - BK)^{-1}}^TK^T.
\end{equation}
We define a new decision variable $L = K(I - BK)^{-1}$\cite{okamoto2018optimal}, such that
\begin{align}
\text{Cov}(\mathbf{X}) &= (I + BL)S(I+BL)^T, \\
\text{Cov}(\mathbf{U}) &= LSL^T, \\
S &= A\Sigma_0A^T + G\Sigma_WG^T.
\end{align}
We consider a feasible control set of the form $\mathcal{U} := \{\mathbf{u}_{\min} \leq \mathbf{u}_k \leq \mathbf{u}_{\max}\}$ for all $k$. We can see that $\mathbf{u}_k \in \mathcal{U} \iff \mathbf{u}_k \leq \mathbf{u}_{\max} \text{ and } -\mathbf{u}_k \leq -\mathbf{u}_{\min}$. Then $\mathcal{U}$ can be represented by the intersection of linear inequality constraints, with $\mathcal{U} := \bigcap_{j=1}^M \{\alpha_j \mathbf{u}_k \leq \beta_j\}$, where $M=2$, $\alpha_1 = 1$, $\alpha_2 = -1$, $\beta_1 = \mathbf{u}_{\max}$, and $\beta_2 = -\mathbf{u}_{\min}$. 

Okamoto et al. \cite{okamoto2018optimal} show that the constraint $\mathbf{u}_k \in \mathcal{U}$ can be reformulated by the set of convex chance constraints $\forall j,\ \mathbb{P}(\alpha_j\mathbf{u}_k \leq \beta_j) \geq 1 - \epsilon_u$. At each time step $k$, $\mathbf{u}_k$ is a Gaussian random variable\cite{okamoto2018optimal, ridderhof2022chance} with mean $E_k^u \overline{\mathbf{U}}$ and covariance $E_k^u \text{Cov}(\mathbf{U}){E_k^u}^T$, where $E_k^u \in \mathbb{R}^{m \times Nm}$ and $E_k^u \mathbf{U} = \mathbf{u}_k$. Then\cite{okamoto2018optimal, ridderhof2022chance},
\begin{equation}
\mathbb{P}(\alpha_j \mathbf{u}_k \leq \beta_j) = \Phi\left(\frac{\beta_j - \alpha_jE_k^u \overline{\mathbf{U}}}{\sqrt{\alpha_j E_k^u \text{Cov}(\mathbf{U}){E_k^u}^T\alpha_j^T}}\right),
\end{equation}
where $\Phi(\cdot)$ is the cumulative distribution function of the standard normal distribution. 

Recalling that $\text{Cov}(\mathbf{U}) = LSL^T$, with $S^{1/2}$ such that $S^{1/2}(S^{1/2})^T = S$ and ${S^T}^{1/2} = (S^{1/2})^T$, the control chance constraints can be reformulated exactly in convex form by \cite{okamoto2018optimal}
\begin{equation}\label{eq: generic_control_constraint}
\beta_j \geq \alpha_jE_k^u \overline{\mathbf{U}} + \Phi^{-1}(1 - \epsilon_u)||{S^T}^{1/2}L^T{E_k^u}^T\alpha_j^T||_2 \quad \forall j.
\end{equation}

With $M=2$, $\alpha_1 = 1$, $\alpha_2 = -1$, $\beta_1 = \mathbf{u}_{\max}$, and $\beta_2 = -\mathbf{u}_{\min}$, we have
\begin{equation}\label{eq: max_bank_angle_constraint}
\mathbf{u}_{\max} \geq E_k^u \overline{\mathbf{U}} + \Phi^{-1}(1 - \epsilon_u)||{S^T}^{1/2}L^T{E_k^u}^T||_2,
\end{equation}
\begin{equation}\label{eq: min_bank_angle_constraint}
-\mathbf{u}_{\min} \geq -E_k^u \overline{\mathbf{U}} + \Phi^{-1}(1 - \epsilon_u)||{S^T}^{1/2}L^T{E_k^u}^T||_2.
\end{equation}
\subsection{Robust Sampling-Based Objective}
For problems with nonlinear objective functions, even if the evolution of the state uncertainty is well-represented by the linear mean and covariance dynamics given in Eq. \ref{eq: mean_state_dynamics} and Eq. \ref{eq: covariance_dynamics}, the objective function may not be well-represented by a Gaussian distribution. We construct a robust sampling-based objective which accounts for variation in the nonlinear objective function by approximating the state trajectory by a collection of sigma point trajectories.

We sample $2n$ sigma points $\mathbf{x}_0^{(1)}, \mathbf{x}^{(i)}_0, \ldots, \mathbf{x}_0^{(2n)}$ symmetrically on the 3rd covariance contour of the initial state distribution, such that
\begin{equation}
\mathbf{x}^{(i)}_0 =
\begin{cases}
\mathbf{x}_0 + 3 {\Sigma_0}_i^{1/2} \qquad i \leq n \\
\mathbf{x}_0 - 3 {\Sigma_0}_i^{1/2} \qquad n < i \leq 2n
\end{cases}
\end{equation}
where ${\Sigma_0}_i^{1/2}$ is the $i$th column of $\Sigma_0^{1/2}$, with $\Sigma_0^{1/2}{\Sigma_0^{1/2}}^T = \Sigma_0$.

We propagate nonlinear sigma point trajectories using the initial control reference trajectory, such that $\forall i$,
\begin{equation}\label{eq: sigma_point_propagation}
\dot{\hat{\mathbf{x}}}^{(i)} = f(\hat{\mathbf{x}}^{(i)}, \hat{\mathbf{u}}, \mathbb{E}[\Psi(\phi(\hat{\mathbf{x}}))])
\end{equation}
Then, we solve a least-squares problem for $\overline{\mathbf{W}}^{(i)}$, a disturbance vector which captures dynamic nonlinearities and the state dependence of the disturbance, such that each nonlinear sigma point trajectory is expressed in block-matrix form by
\begin{equation}\label{eq: sigma_point_mean_dynamics}
\hat{\mathbf{X}}^{(i)} = A\mathbf{x}_0^{(i)} + B\overline{\mathbf{U}}+ C + G\overline{\mathbf{W}}^{(i)}
\end{equation}
with $A, B, C, G$ linearized around the system mean trajectory, and with $\hat{\mathbf{X}}^{(i)}$ equal to stacking $\hat{\mathbf{x}}(t_k)$ for all $k$.

Next, we evaluate and linearize the nonlinear objective function around each sigma point trajectory. Recall from Problem \ref{prob: aerocapture_delta_V} that the nonlinear objective function is given by
\begin{equation}
    \inf \{\gamma \in \mathbb{R}: \mathbb{P}(\Delta V(\mathbf{x}_N) > \gamma) \leq X/100\},
\end{equation}
where $\mathbf{x}_N$ is the final trajectory state at time $t_N$. We represent $\Delta V(\mathbf{x}_N)$ by a first-order Taylor approximation with
\begin{equation}
\Delta V(\mathbf{x}_N) \approx \Delta V(\hat{\mathbf{x}}_N) + \frac{\partial \Delta V}{\partial \mathbf{x}_N}|_{\hat{x}_N} (\mathbf{x}_N - \hat{\mathbf{x}_N}).
\end{equation}
Then, 
\begin{equation}
\mathbb{P}(\Delta V(\mathbf{x}_N) \leq \gamma) \approx \mathbb{P}\left(\frac{\partial \Delta V}{\partial \mathbf{x}_N}|_{\hat{x}_N} \mathbf{x}_N \leq \gamma - \Delta(\hat{\mathbf{x}_N}) + \frac{\partial \Delta V}{\partial \mathbf{x}_N}|_{\hat{x}_N}  \hat{\mathbf{x}_N}\right)
\end{equation}
Defining $\xi = \gamma - \Delta V(\hat{\mathbf{x}_N}) + \frac{\partial \Delta V}{\partial \mathbf{x}_N}|_{\hat{x}_N}  \hat{\mathbf{x}_N}$ as an auxiliary variable, we have
\begin{equation}
\mathbb{P}\left(\frac{\partial \Delta V}{\partial \mathbf{x}_N}|_{\hat{x}_N}\mathbf{x}_N \leq \xi\right) = \Phi^{-1}\left(\frac{\xi - \frac{\partial \Delta V}{\partial \mathbf{x}_N}|_{\hat{x}_N}\mathbb{E}[\mathbf{x}_N]}{\sqrt{\frac{\partial \Delta V}{\partial \mathbf{x}_N}|_{\hat{x}_N}\Sigma_N \frac{\partial \Delta V}{\partial \mathbf{x}_N}|_{\hat{x}_N}^T}}\right)
\end{equation}
where $\Sigma_N$ is the state covariance at time $t_N$, and that
\begin{eqnarray} \hspace*{-0.25in}
\mathbb{P}\left(\frac{\partial \Delta V}{\partial \mathbf{x}_N}|_{\hat{x}_N} \mathbf{x}_N > \xi\right) \leq X/100 
&\!\!\iff\!\!& \frac{\partial \Delta V}{\partial \mathbf{x}_N}|_{\hat{x}_N}\mathbb{E}[\mathbf{x}_N] + \notag \\ 
&& \Phi^{-1}(1- X/100)\sqrt{\frac{\partial \Delta V}{\partial \mathbf{x}_N}|_{\hat{x}_N}\Sigma_N \frac{\partial \Delta V}{\partial \mathbf{x}_N}|_{\hat{x}_N}^T} \leq \xi
\end{eqnarray}
It follows that
\begin{align}
\inf &\{\gamma \in \mathbb{R}: \mathbb{P}(\Delta V(\mathbf{x}_N) > \gamma) \leq X/100\} \approx \notag \\
&\min \Delta V(\hat{\mathbf{x}}_N) + \frac{\partial \Delta V}{\partial \mathbf{x}_N}|_{\hat{x}_N}(\mathbb{E}[\mathbf{x}_N] - \hat{\mathbf{x}}_N) + \Phi^{-1}(1- X/100)\sqrt{\frac{\partial \Delta V}{\partial \mathbf{x}_N}|_{\hat{x}_N}\Sigma_N \frac{\partial \Delta V}{\partial \mathbf{x}_N}|_{\hat{x}_N}^T}
\end{align}
However, because $\Delta V$ is a nonlinear function of the final state, even if the final state is well-approximated by a Gaussian distribution, $\Delta V(\mathbf{x}_N)$ may vary significantly over the final state distribution. As such, we sample values of the objective at different points in the final state distribution, as specified by the sigma point trajectories given by Eq. \ref{eq: sigma_point_mean_dynamics}. Each sigma point trajectory has a different $\hat{\mathbf{x}}_N^{(i)}$, and also has a different mean state given by
\begin{align}
\mathbb{E}[\mathbf{x}_N^{(i)}] &= E_N^x\hat{\mathbf{X}}^{(i)} = E_N^x(A\mathbf{x}_0^{(i)} + B \overline{\mathbf{U}} + \mathbf{C} + G\overline{\mathbf{W}}^{(i)}).
\end{align}
For each sampled final state distribution, we use the linear approximation of the nominal covariance dynamics given in Eq. \ref{eq: covariance_dynamics},
such that
\begin{align}
\Sigma_N^{(i)} = \Sigma_N = E_N^x \text{Cov}(\mathbf{X})E_N^x = E_N^x(I + BL)S(I+BL)^T{E_N^x}^T,
\end{align}
in order to penalize growth in the final state covariance. Our robust sampling-based cost function is given by
\begin{align}
\max_i \Delta V\left(\hat{\mathbf{x}}_N^{(i)}\right) &+ \frac{\partial \Delta V}{\partial \mathbf{x}_f}|_{\hat{\mathbf{x}}_N^{(i)}}\left(E_N^x\hat{\mathbf{X}}^{(i)}- \hat{\mathbf{x}}_f^{(i)}\right) \notag \\
&+ \Phi^{-1}(1- X/100)\sqrt{\frac{\partial \Delta V}{\partial \mathbf{x}_N}|_{\hat{\mathbf{x}}_N^{(i)}}E_N^x \text{Cov}(\mathbf{X}){E_N^x}^T \frac{\partial \Delta V}{\partial \mathbf{x}_N}|_{\hat{\mathbf{x}}_N^{(i)}}^T},
\end{align}
which is approximately equal to the worst-case $X$th-percentile $\Delta V$ over a collection of Gaussian trajectories, where the means of the Gaussians are given by the sigma point trajectories and the covariance of each Gaussian is given by the system covariance dynamics. 

Recalling that $\text{Cov}(\mathbf{X}) = (I+BL)S(I+BL)^T$, this function can be rewritten in convex form as
\begin{align}\label{eq: robust_objective}
\max_i \Delta V\left(\hat{\mathbf{x}}_N^{(i)}\right) &+ \frac{\partial \Delta V}{\partial \mathbf{x}_N}|_{\hat{\mathbf{x}}_N^{(i)}}\left(E_N^x\hat{\mathbf{X}}^{(i)}- \hat{\mathbf{x}}_N^{(i)}\right) \notag \\ 
&+ \Phi^{-1}(1- X/100)\left|\left|{S^T}^{1/2}(I + BL)^T{E_N^x}^T \frac{\partial \Delta V}{\partial \mathbf{x}_N}|_{\hat{\mathbf{x}}_N^{(i)}}^T\right|\right|_2.
\end{align}

Our cost function captures variation in $\Delta V(\mathbf{x}_N)$ and $\frac{\partial \Delta V}{\partial \mathbf{x}_N}$ over the state distribution, penalizing sigma point trajectories associated with high $\Delta V$, and penalizing changes in the control plan which would increase $\Delta V$ for the worst-case sigma point trajectory. Additionally, propagating the nominal sigma point trajectories $\hat{\mathbf{x}}^{(0)}, \ldots, \hat{\mathbf{x}}^{(2n)}$ using the true system dynamics captures variation in $\Delta V(\mathbf{x}_N)$ resulting from dynamic nonlinearities and nonlinearities caused by the state-dependence of the atmosphere. Finally, using the nominal final state covariance for each sigma point increases robustness and penalizes growth in the final state covariance.

\subsection{Robust Sampling-based Covariance Steering via Successive Convexification}
With the control constraints given in Eq. \ref{eq: max_bank_angle_constraint}-\ref{eq: min_bank_angle_constraint} and our robust sampling-based objective given in Eq. \ref{eq: robust_objective}, we can construct a convex problem which approximates Problem \ref{prob: aerocapture_delta_V}. However, we must introduce trust region constraints to ensure that the convex approximation of the nonlinear problem remains valid. Following Ridderhof \& Tsiotras \cite{ridderhof2022chance}, we construct state and control trust region constraints of the form
\begin{equation}
||\overline{\mathbf{u}}_k - \hat{\mathbf{u}}_k||_{M_k^u} \leq \Delta^u,
\end{equation}
\begin{equation}
||\overline{\mathbf{x}}_k - \hat{\mathbf{x}}_k||_{M_k^x} \leq \Delta^x,
\end{equation}
where $\Delta^u, \Delta^x \in \mathbb{R}_{\geq 0}$ and $M_k^u, M_k^x \in S_m^+, S_n^+$. 

Specifically, we select $\Delta^u = 0.1$ and $M_k^u = I$ to keep $\overline{\mathbf{U}}$ close to the nominal control trajectory\cite{ridderhof2022chance}. For non-terminal states, we construct our state trust region constraints to penalize deviations in the dynamic pressure $q = \rho(r)v^2/2$, as large deviations in $q$ represent large deviations from the nominal lift and drag forces, invalidating the linear approximation of the dynamics \cite{ridderhof2022chance, albert2023onboard}. For the terminal state, we penalize deviations in the vehicle exit apoapsis $r_{a, \text{exit}}$, as excessive deviations in the vehicle exit apoapsis will invalidate our robust sample-based approximation of the $\Delta V$ required for aerocapture. As such, for $k < N$, we select
\begin{equation}
M_k^x = \hat{q}^{-2}\left(\frac{\partial \hat{q}}{\partial \hat{x}_k} \right)^T\left(\frac{\partial \hat{q}}{\partial \hat{x}_k} \right),\quad \Delta^x = 10^{-3}
\end{equation}
and when $k = N$, as in Ridderhof \& Tsiotras\cite{ridderhof2022chance}, we select
\begin{equation}
M_N^x = \left(\frac{\partial r_{a, \text{exit}}}{\partial x} \right)^T\left(\frac{\partial r_{a, \text{exit}}}{\partial x} \right), \quad \Delta^x = 0.1 r_p,
\end{equation}
where $r_p$ is the radius of the planet. We formalize a trust-constrained convex robust sampling-based approximation of Problem \ref{prob: aerocapture_delta_V} in Problem \ref{prob: convex_subproblem}.
\begin{problem} \label{prob: convex_subproblem}
\textit{Minimize the Xth-percentile $\Delta V$ required for successful aerocapture over a collection of Gaussian trajectories, subject to probabilistic bank angle constraints and state and control trust region constraints.}
\begin{align}
\min_{L, \overline{\mathbf{U}}} \max_i \Delta V\left(\hat{\mathbf{x}}_N^{(i)}\right) &+ \frac{\partial \Delta V}{\partial \mathbf{x}_N}|_{\hat{\mathbf{x}}_N^{(i)}}\left(E_N^x\left(A\mathbf{x}_0^{(i)} + B \overline{\mathbf{U}} + \mathbf{C} + G\overline{\mathbf{W}}^{(i)}\right)- \hat{\mathbf{x}}_N^{(i)}\right) \notag \\ 
&+ \Phi^{-1}(1- X/100)\left|\left|{S^T}^{1/2}(I + BL)^T{E_N^x}^T \frac{\partial \Delta V}{\partial \mathbf{x}_N}|_{\hat{\mathbf{x}}_N^{(i)}}^T\right|\right|_2
\end{align}
such that:
\begin{equation}
\begin{split}
\hat{\mathbf{X}}^{(i)} &= A\mathbf{x}_0^{(i)} + B\overline{\mathbf{U}}+ C + G\overline{\mathbf{W}}^{(i)}\ \forall i\\
S &= A\Sigma_0A^T + G\Sigma_WG^T \\
 -\mathbf{u}_{\min} &\geq -E_k^u \overline{\mathbf{U}} + \Phi^{-1}(1 - \epsilon_u)||{S^T}^{1/2}L^T{E_k^u}^T||_2\ \forall k \\
 \mathbf{u}_{\max} &\geq E_k^u \overline{\mathbf{U}} + \Phi^{-1}(1 - \epsilon_u)||{S^T}^{1/2}L^T{E_k^u}^T||_2\ \forall k \\
 ||E_k^u \overline{\mathbf{U}} - \hat{\mathbf{u}}(t_k)||_{M_k^u} & \leq \Delta_u\ \forall k \\
 ||E_k^x \overline{\mathbf{X}} - \hat{\mathbf{x}}(t_k)||_{M_k^x} & \leq \Delta_u\ \forall k
\end{split}
\end{equation}
\end{problem}
Because Problem \ref{prob: convex_subproblem} only approximates Problem \ref{prob: aerocapture_delta_V} within a narrow region specified by the trust constraints, we use successive convex programming, repeatedly solving Problem \ref{prob: convex_subproblem} and re-linearizing the system after each iteration. Our successive convex programming algorithm is given in Algorithm~\ref{alg: successive_convexification}.

\begin{algorithm}[t]\label{alg: successive_convexification}
\caption{Iterative robust sampling-based covariance steering in a Gaussian random field}
\textbf{Input:} {Nominal control trajectory $\hat{\mathbf{u}}$, initial state mean $\overline{\mathbf{x}}_0$, initial state covariance $\Sigma_0$, time discretization $[t_0, \ldots t_N]$}

\textbf{Output:} {Nominal control trajectory $\overline{\mathbf{U}}$, feedback control gain $K$, nominal state trajectory $\overline{\mathbf{X}}$}

\While{termination criteria not met}{
Propagate nominal trajectory (Eq. \ref{eq: nominal_trajectory_propagation}); \\
Linearize and discretize around the nominal trajectory (Eqs. \ref{eq: discrete_time_dynamics}-\ref{eq: block_matrix_dynamics}); \\
Propagate sigma point trajectories (Eq. \ref{eq: sigma_point_propagation}); \\
Linearize and discretize sigma point trajectories (Eq. \ref{eq: sigma_point_mean_dynamics}); \\
Solve Problem \ref{prob: convex_subproblem} to find $L, \overline{\mathbf{U}}$; \\
Update feedback control gain: $K \gets L(I+BL)^{-1}$; \\
Update nominal control: $\hat{\mathbf{u}}_k \gets \overline{\mathbf{u}}_k \ \forall k$; \\
}
\end{algorithm}
\section{Results}
This section presents the results of closed-loop aerocapture simulations at Mars and Uranus. We run a state-of-the-art covariance steering baseline algorithm \cite{ridderhof2022chance} and our proposed robust sampling-based covariance steering algorithm offline with a known initial state distribution and distribution over atmospheric density perturbations, generating a nominal control trajectory $\overline{\mathbf{U}}$ and a feedback control gain matrix $K$. Then, we simulate Monte Carlo trajectories with dispersed initial state and atmospheric conditions, and run the closed-loop control found by each guidance algorithm, clipping all control inputs to stay within specified control bounds. We evaluate the performance of both methods in terms of final $\Delta$V required for successful aerocapture.

\subsection{Aerocapture at Mars}
We present results for two Mars aerocapture scenarios. Both Mars aerocapture experiments use the same initial state, target conditions, initial control guess, and vehicle parameters as Ridderhof \& Tsiotras \cite{ridderhof2022chance}. However, unlike in prior work \cite{ridderhof2022chance}, we include initial state uncertainty in both scenarios, increasing the difficulty and realism of each aerocapture scenario. The initial state and vehicle parameters for both scenarios are presented in Table \ref{tab:mars_parameters}. 
\begin{table}[htbp]
	\fontsize{10}{10}\selectfont
    \caption{Initial and target conditions and vehicle parameters for Mars aerocapture.}
    \label{tab:mars_parameters}
        \centering 
   \begin{tabular}{|c|c|} 
      \hline 
      \textbf{Parameter} & \textbf{Value}\\
      \hline 
      Initial planet-relative altitude, $h_0$ (km) & 125 \\
      Initial planet-relative velocity, $v_0$ (km/s) & 6.1 \\
      Initial flight-path angle, $\gamma_0$ (deg) & -10.0128 \\
      Ballistic coefficient, $B_c$ (kg/m$^2$) & 150 \\
      Nominal L/D ratio & 0.2 \\
      Target apoapsis, $r_a^*$ (km) & 16985 \\
      Target periapsis, $r_p^*$ (km) & 6794\\
      Maximum cosine bank angle, $u_{\max}$ & 1 \\
      Minimum cosine bank angle, $u_{\min}$ & -1 \\
      \hline
   \end{tabular}
\end{table}

Both experiments also use the same initial control guess, with $\hat{\mathbf{u}}_k = 0$ for all $k$, and use the same atmospheric model, with atmospheric density
$\rho = \overline{\rho}(1 + \delta p/100)$,
where $\overline{\rho}$ is given by the nominal MarsGRAM atmospheric density, and $\delta p$ is a zero-mean Gaussian random field with covariance function
\begin{equation}
\Sigma(h_1, h_2) = \exp\left(-\frac{|h_1 - h_2|}{H_\text{scale}} \right) \times \begin{cases} b(\min(h_1, h_2)), \quad \min(h_1, h_2) < h_\text{trans} \\ \sigma^2_{\rho, \max}, \qquad \qquad \ \ \min(h_1, h_2) \geq h_\text{trans}\end{cases}
\end{equation}
where $H_\text{scale} = 11.1$ km is the atmosphere scale height, and
\begin{equation}
b(h) = \sigma^2_{\rho, \max}\exp\left(\frac{h - h_\text{trans}}{c_\text{scale}}\right)
\end{equation}
with $c_\text{scale} = 20$ km, $h_\text{trans} = 120$ km, and the maximum density variance $\sigma^2_{\rho, \max} = 1480$ (kg/m$^3$)$^2$.

\begin{figure}[t]
	\centering\includegraphics[width=2.5in]{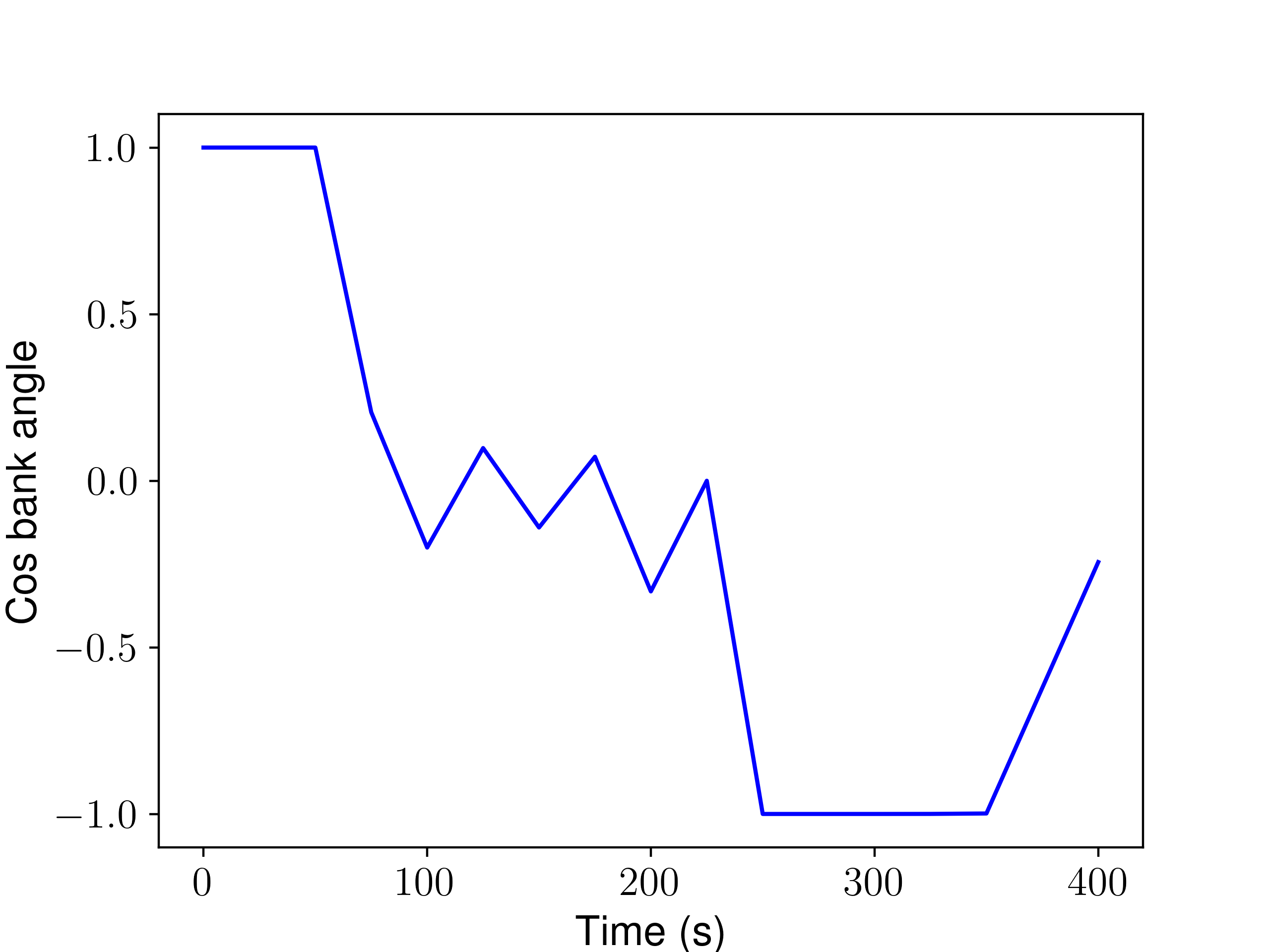}
    \includegraphics[width=2.5in]{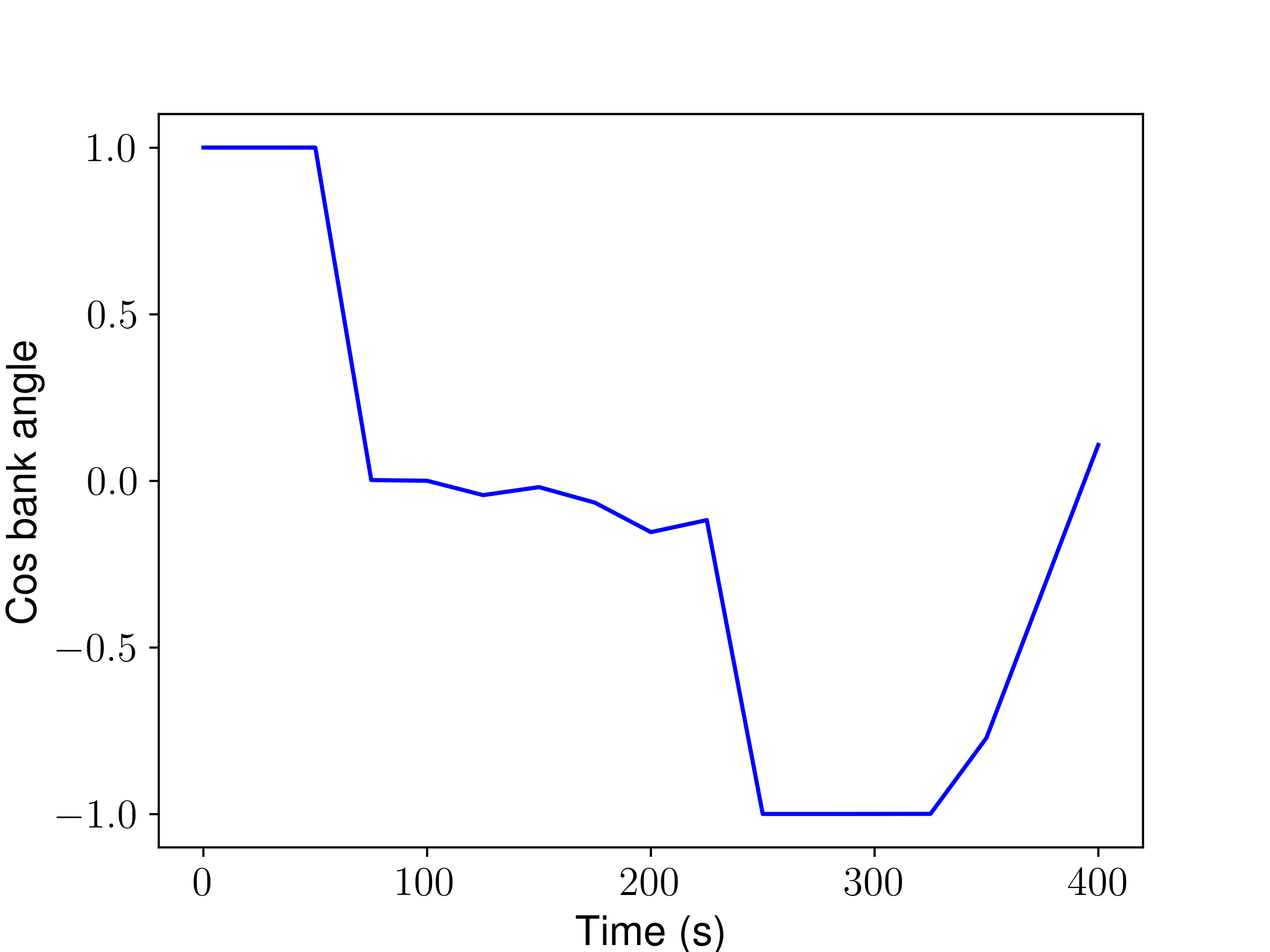}
	\caption{Nominal control trajectory after 30 iterations for our method for a Mars aerocapture case with a small initial state dispersion (left) and a Mars aerocapture case with a large initial velocity dispersion (right).}
	\label{fig:mars_nominal_control}
\end{figure}

We use the same time discretization as Ridderhof \& Tsiotras, except that we add one additional time step, setting the final time to $t_f = 450$ sec in order to allow slow entry cases to exit the atmosphere by the final time. Our time discretization is given by $[0, 50, 75, 100, 125, 150, 175, 200, 225,$ $ 250, 275, 300, 325, 350, 400, 450]$ seconds. When implementing the baseline algorithm \cite{ridderhof2022chance}, we also modify the objective to minimize 99th-percentile $\Delta V$, enforcing state trust region constraints on the dynamic pressure rather than including a secondary objective to minimize deviation in dynamic pressure. We run the iterative robust sampling-based covariance steering algorithm given in Algorithm \ref{alg: successive_convexification} for 30 iterations in order to allow the nominal control trajectory to fully converge. 

For both of our Mars experiments, nominal control trajectories after 30 iterations resemble bang-zero-bang trajectories, as shown in Figure \ref{fig:mars_nominal_control}. For the deterministic aerocapture problem, when the initial state and atmosphere are perfectly known, the optimal control trajectory has a bang-bang structure \cite{lu2015optimal}. However, when the atmospheric density and initial state are dispersed, the control switching time varies. Intuitively, a bang-zero-bang nominal trajectory  allows for all stochastic trajectories to follow a bang-bang structure, but with the switching time determined by the feedback control gain, varying with the atmospheric density and entry state.

\begin{figure}[t]
	\centering\includegraphics[width=2.5in,trim=0 0 0 20,clip]{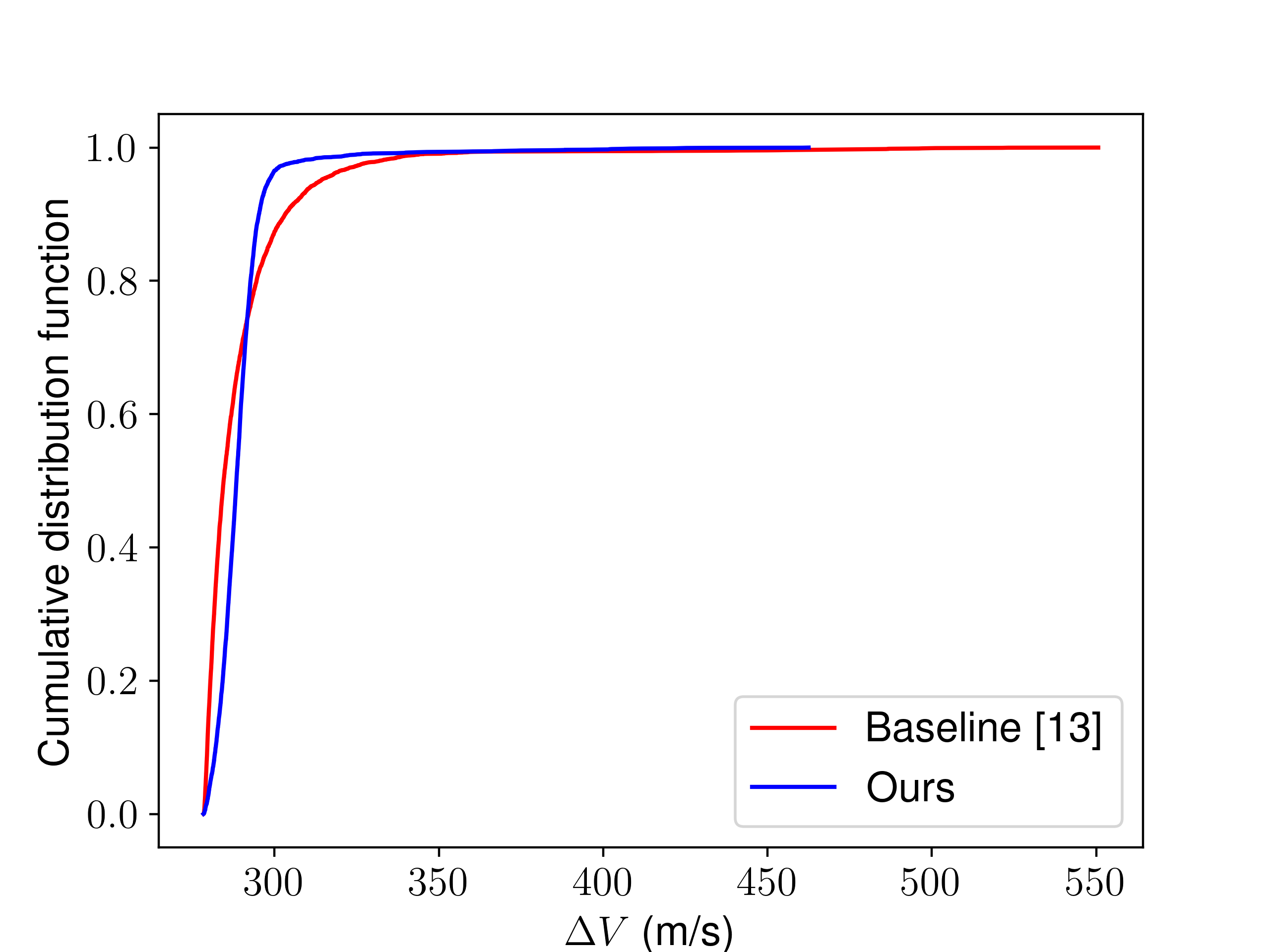}
    \includegraphics[width=2.5in,trim=0 0 0 20,clip]{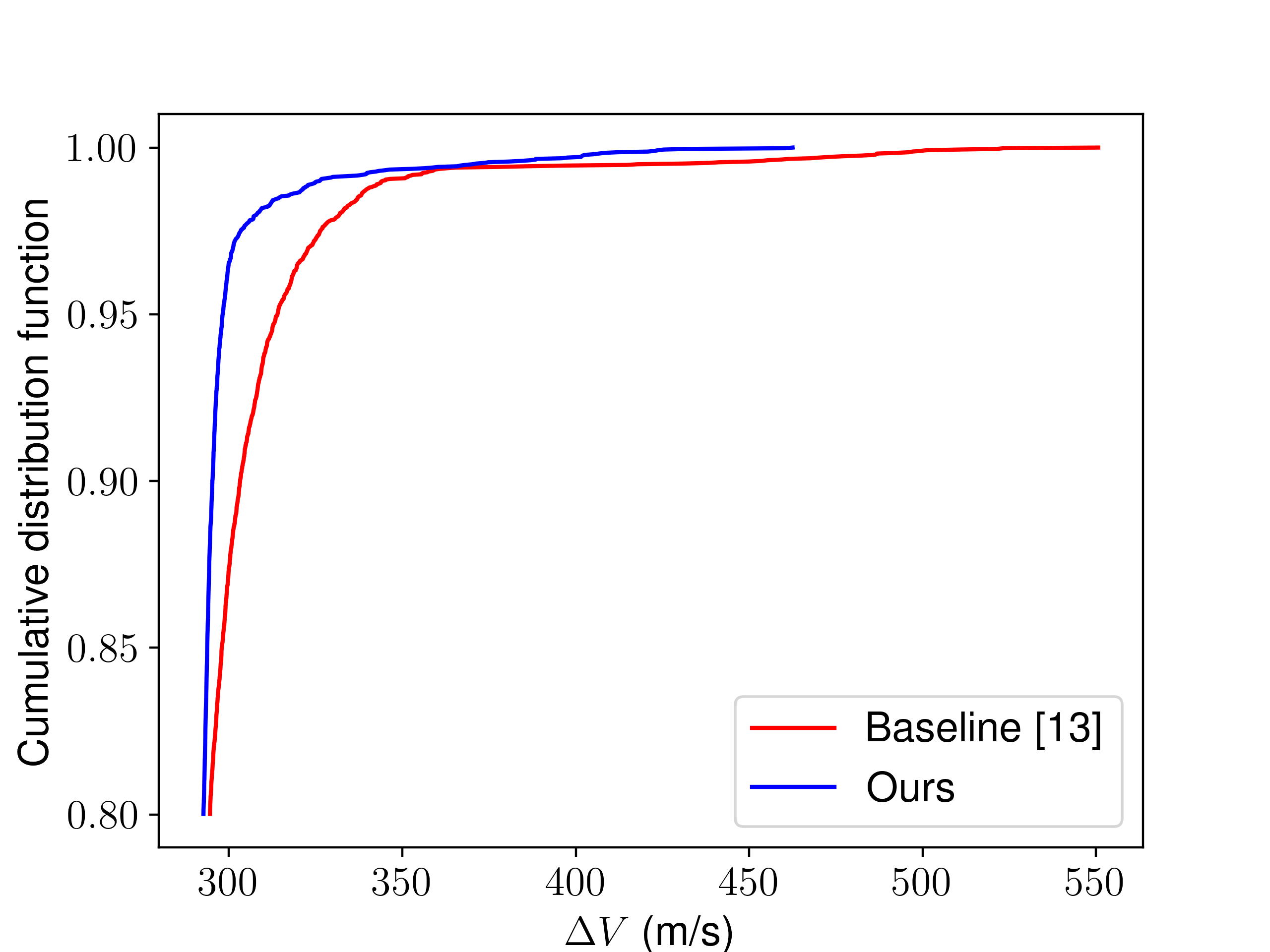}
	\caption{Full $\Delta V$ cumulative distribution function for our method and the baseline \cite{ridderhof2022chance} with a small initial state dispersion (left) and zoomed-in $\Delta V$ cumulative distribution function with a small initial state dispersion (right).}
	\label{fig:mars_easy_cdf}
	\centering\includegraphics[width=2.5in,trim=0 0 0 20,clip]{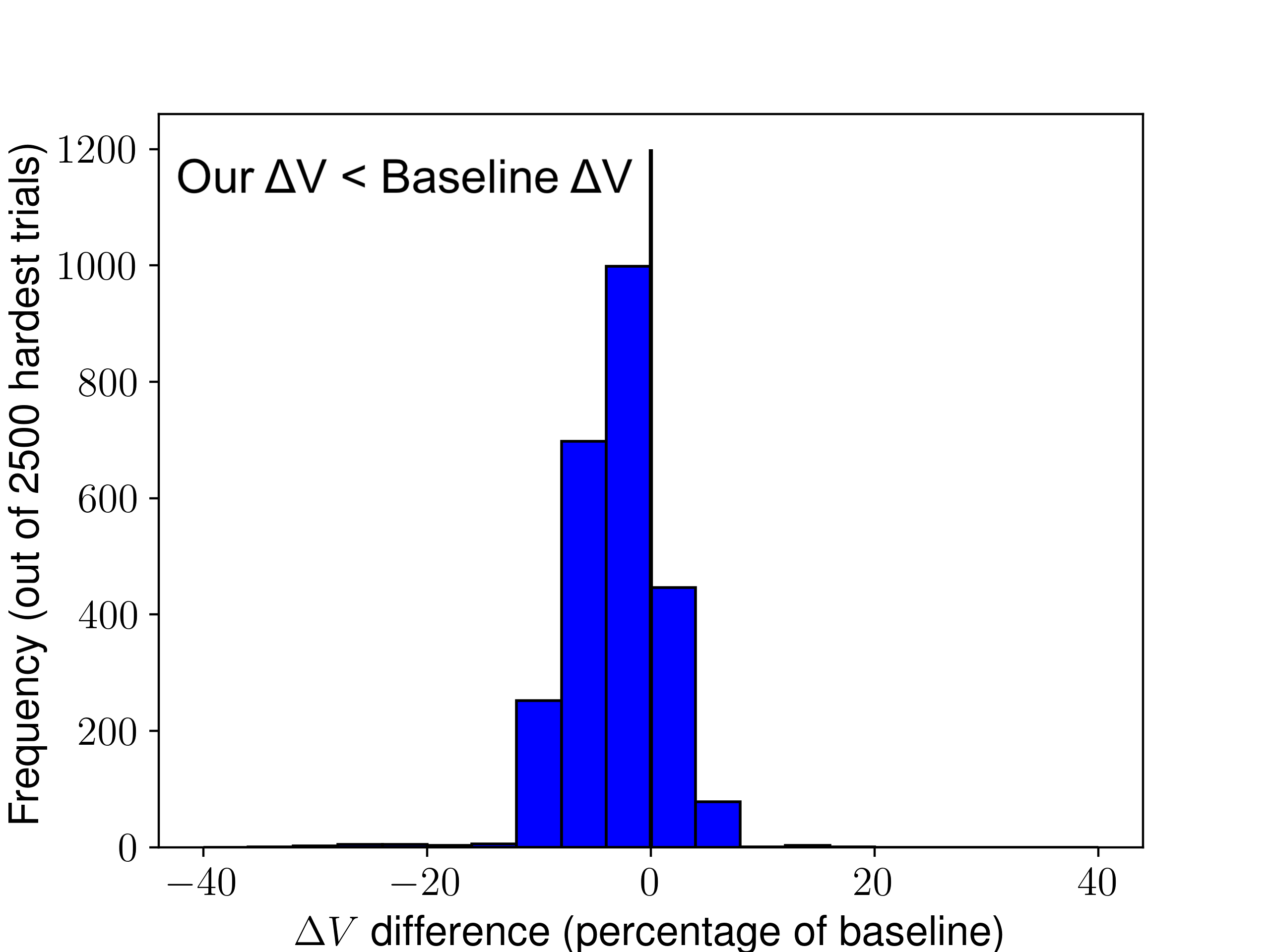}
    \includegraphics[width=2.5in,trim=0 0 0 20,clip]{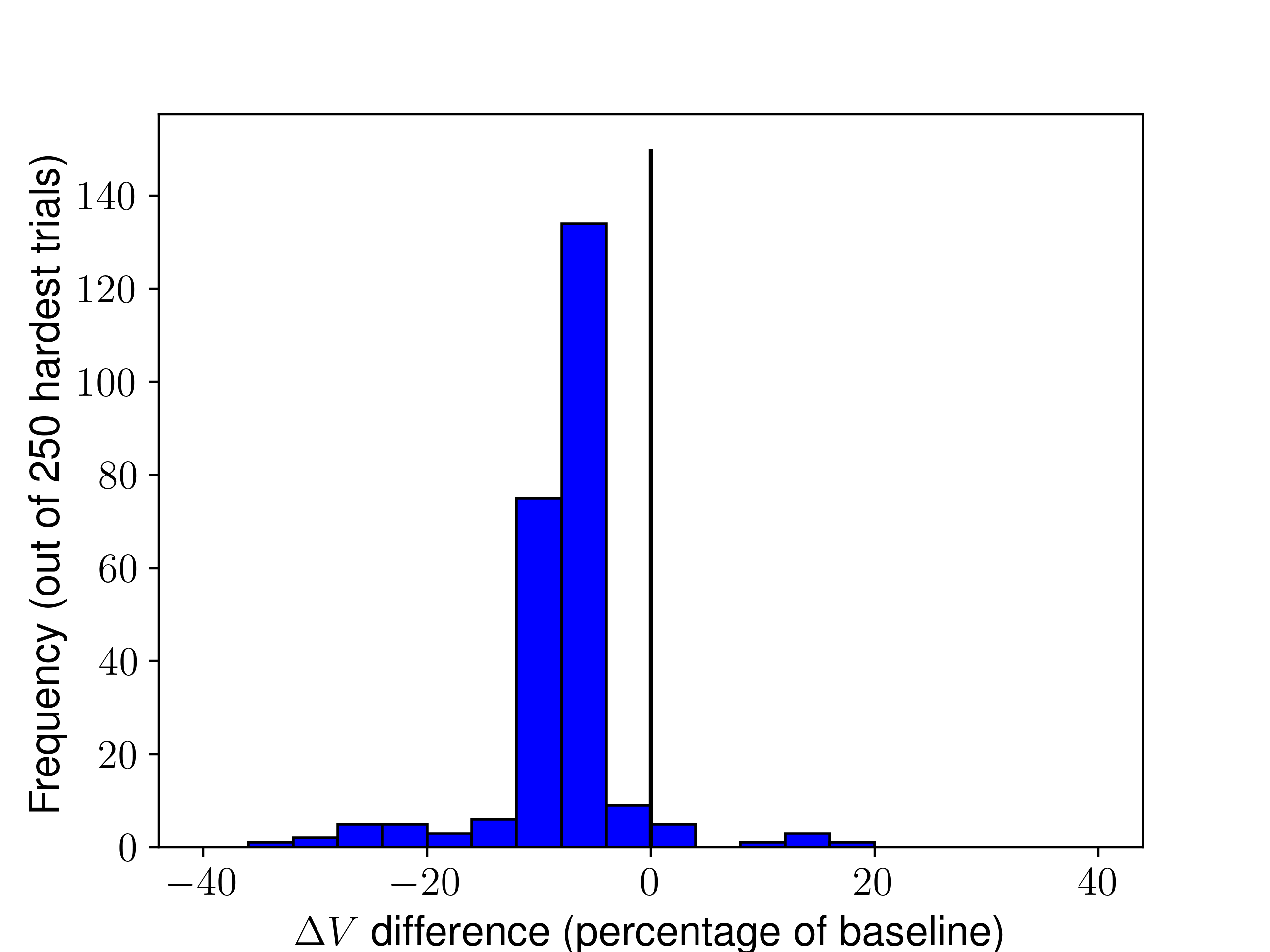}
	\caption{Difference between final $\Delta V$ for our method and the baseline \cite{ridderhof2022chance} for the hardest 2500 Mars aerocapture cases with a small initial state dispersion (left) and the hardest 250 Mars aerocapture cases with a small initial state dispersion (right).}
	\label{fig:mars_easy_rel_hist}
\end{figure}

In our first Mars experiment, we use a small initial state dispersion, with the $3\sigma$ initial state dispersion equal to $1$ km in altitude, $0.1$ km/s in the magnitude of the velocity, and $0.1$ degrees in the flight path angle of the velocity vector. We solve for a nominal control trajectory and control feedback gain with our robust sampling-based guidance algorithm, and with the baseline iterative nonlinear covariance steering algorithm \cite{ridderhof2022chance}. We simulate 5000 Monte Carlo trajectories with dispersed initial state and atmospheric conditions, running the closed-loop control found by each algorithm and clipping the control to remain within the bounds specified in Table \ref{tab:mars_parameters}. Our Monte Carlo results are presented in Figures \ref{fig:mars_easy_cdf} and \ref{fig:mars_easy_rel_hist} and Table \ref{tab:mars_easy_results}.

\begin{table}[b]
	\fontsize{10}{10}\selectfont
    \caption{Median, 99th percentile, 99.7th percentile, and maximum $\Delta$V for Mars aerocapture with a small initial state dispersion.}
    \label{tab:mars_easy_results}
        \centering 
   \begin{tabular}{|c|c|c|c|c|c|} 
      \hline 
      \textbf{Method}    & \textbf{Median} $\Delta$V & \textbf{Mean} $\Delta V$ & \textbf{99th pct.} $\Delta$V & \textbf{99.7th pct.} $\Delta$V & \textbf{Maximum} $\Delta$V\\
      \hline 
      Baseline \cite{ridderhof2022chance} & \textbf{284.5} & 289.6 & 344.3 & 470.2 & 550.6 \\
      Ours & 288.4 & \textbf{289.6} & \textbf{326.2} & \textbf{397.6} & \textbf{462.5} \\
      \hline
   \end{tabular}
\end{table}

We find that our robust sampling-based covariance steering objective reduces the 99.7th-percentile and worst-case $\Delta V$ required for aerocapture over 5000 trajectories by 15\%, and reduces the 99th-percentile $\Delta V$ (our objective) by 5\%. Figure \ref{fig:mars_easy_cdf} shows the $\Delta V$ cumulative distribution function for our method and for the baseline. Figure \ref{fig:mars_easy_rel_hist} illustrates the $\Delta V$ reduction achieved by our method on the hardest 2500 Monte Carlo cases and on the hardest 500 Monte Carlo cases, demonstrating the benefits of the robust sampling-based covariance steering objective on the most difficult aerocapture scenarios.

\begin{figure}[t]
	\centering\includegraphics[width=2.5in,trim=0 0 0 20,clip]{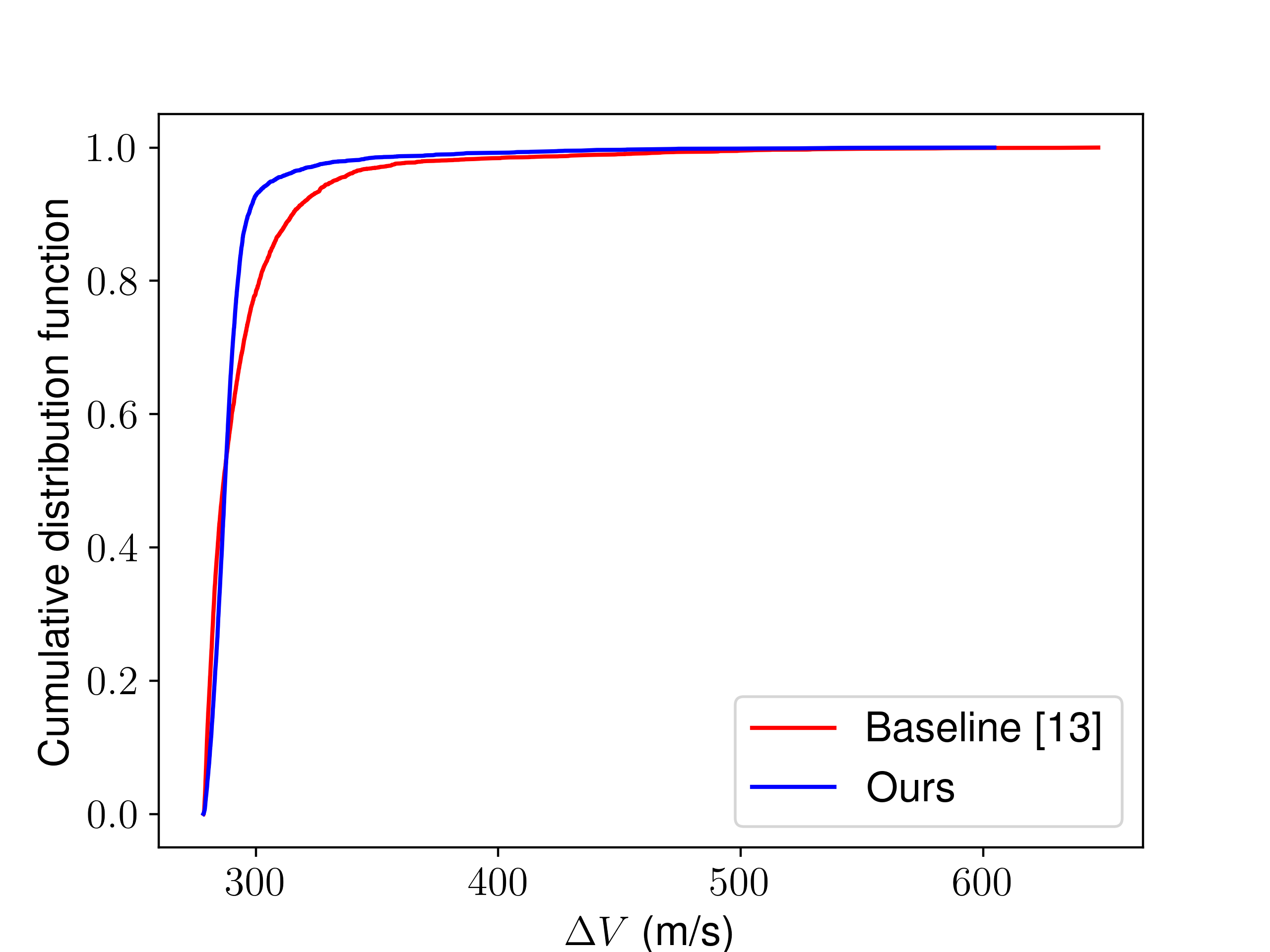}
    \includegraphics[width=2.5in,trim=0 0 0 20,clip]{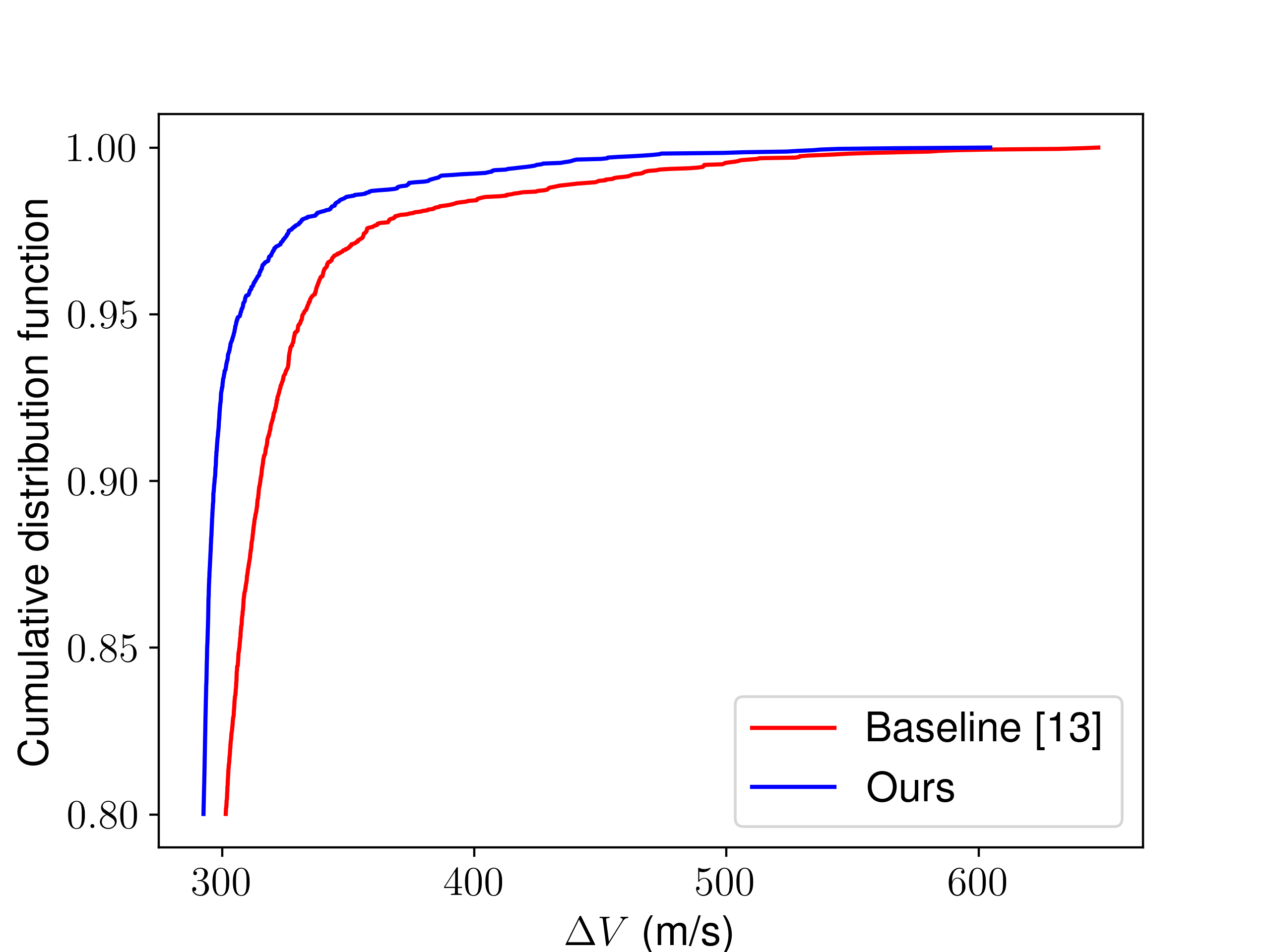}
	\caption{Full $\Delta V$ cumulative distribution function for our method and the baseline \cite{ridderhof2022chance} with a large initial velocity dispersion (left) and zoomed-in $\Delta V$ cumulative distribution function with a large initial velocity dispersion (right).}
	\label{fig:mars_hard_cdf}
\vspace*{-6pt}
	\centering\includegraphics[width=2.5in,trim=0 0 0 20,clip]{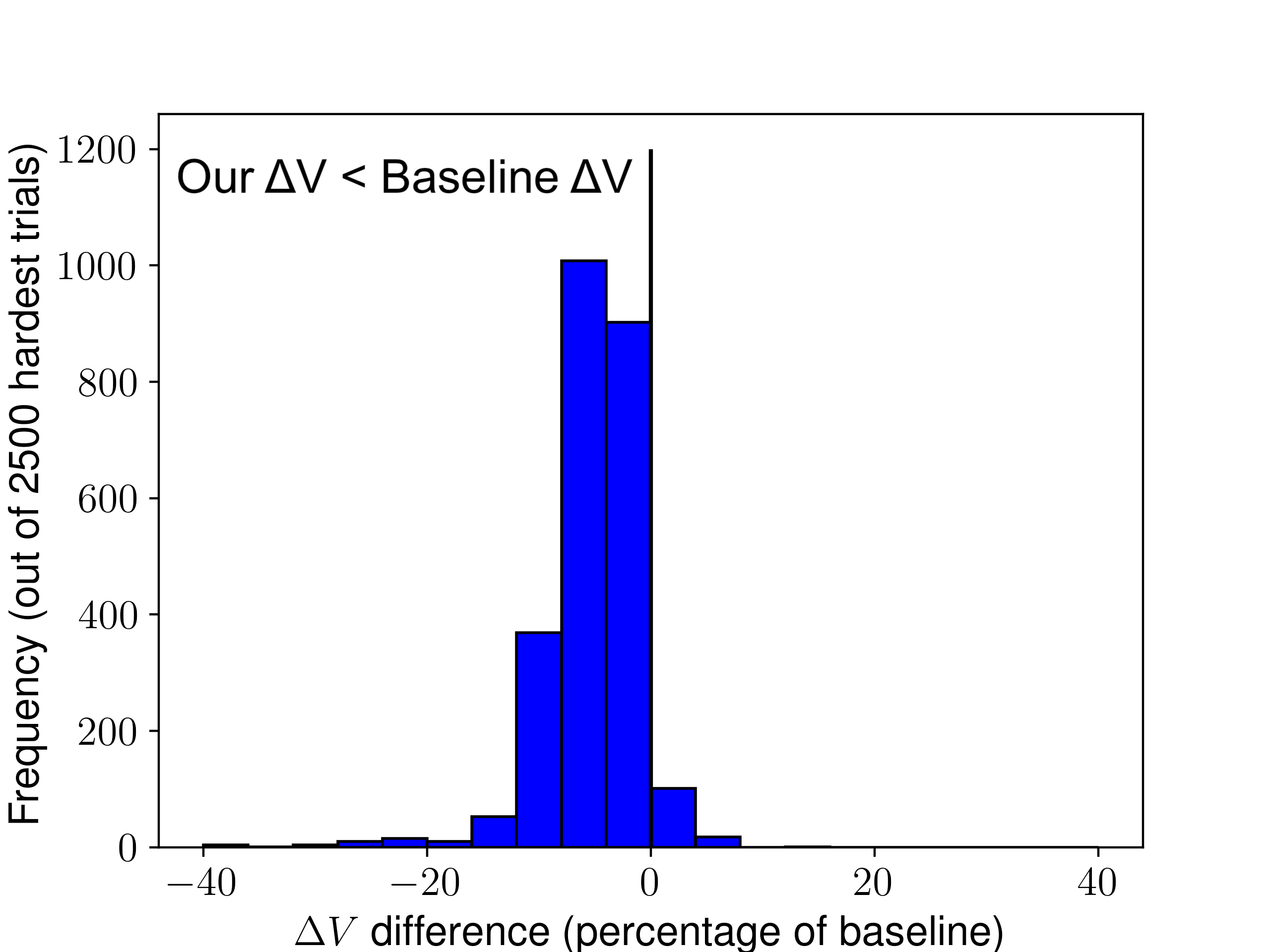}
    \includegraphics[width=2.5in,trim=0 0 0 20,clip]{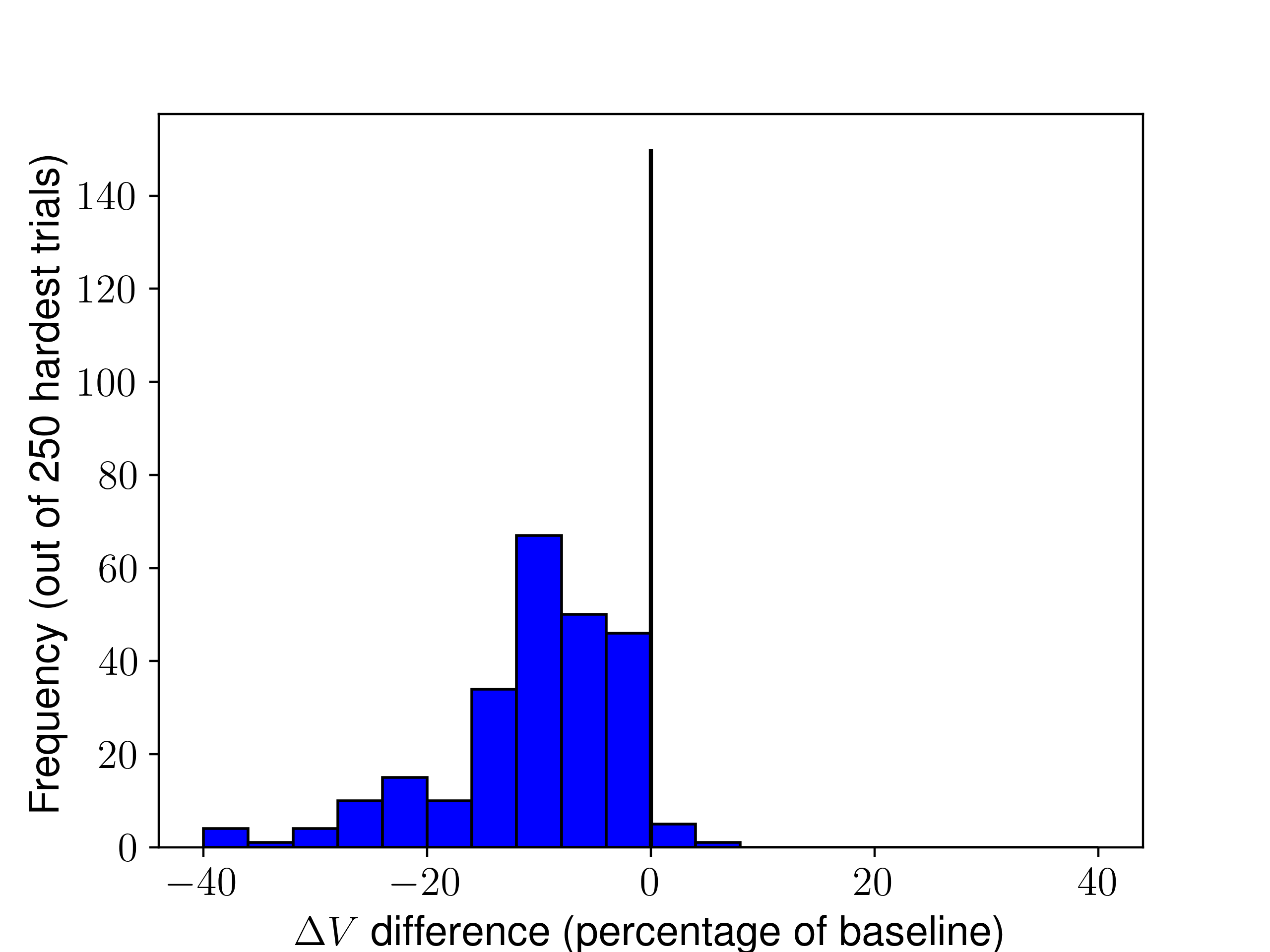}
	\caption{Difference between final $\Delta V$ for our method and the baseline \cite{ridderhof2022chance} for the hardest 2500 Mars aerocapture cases with a large initial velocity dispersion (left) and the hardest 250 Mars aerocapture cases with a large initial velocity dispersion (right).}
	\label{fig:mars_hard_rel_hist}
\end{figure}

We also perform a more challenging Mars aerocapture experiment with a large initial velocity dispersion. In this case, the $3\sigma$ initial state dispersion is equal to $1$ km in altitude, $0.3$ km/s in the magnitude of the velocity, and $0.1$ degrees in the flight path angle of the velocity vector. We solve for a nominal control trajectory and control feedback gain with our robust sampling-based guidance algorithm, and with the iterative nonlinear covariance steering algorithm presented in Ridderhof \& Tsiotras \cite{ridderhof2022chance}. We simulate 5000 Monte Carlo trajectories with dispersed initial state and atmospheric conditions, running the closed-loop control found by each algorithm and clipping the control to remain within the bounds specified in Table \ref{tab:mars_parameters}. Our Monte Carlo results are presented in Figures \ref{fig:mars_hard_cdf} and \ref{fig:mars_hard_rel_hist} and Table \ref{tab:mars_hard_results}.

\begin{table}[b]
	\fontsize{10}{10}\selectfont
    \caption{Median, 99th percentile, 99.7th percentile, and maximum $\Delta$V for Mars aerocapture with a large initial velocity dispersion.}
    \label{tab:mars_hard_results}
        \centering 
   \begin{tabular}{|c|c|c|c|c|c|} 
      \hline 
      \textbf{Method}    & \textbf{Median} $\Delta$V & \textbf{Mean} $\Delta V$ & \textbf{99th pct.} $\Delta$V & \textbf{99.7th pct.} $\Delta$V &\textbf{Maximum} $\Delta$V\\
      \hline 
      Baseline \cite{ridderhof2022chance} & \textbf{286.7} & 295.2 & 449.3 & 527.4 & 647.5 \\
      Ours & 287.3 & \textbf{290.7} & \textbf{381.8} & \textbf{453.0} & \textbf{604.6} \\
      \hline
   \end{tabular}
\end{table}

We find that our robust sampling-based covariance steering objective reduces the worst-case $\Delta V$ required for aerocapture over 5000 trajectories by 6.5\%, reduces the 99.7th-percentile $\Delta V$ by 14\%, and reduces the 99th-percentile $\Delta V$ (our objective) by 15\%. We also see a 1.5\% reduction in mean $\Delta V$. Figure \ref{fig:mars_hard_cdf} shows the $\Delta V$ cumulative distribution function for our method and for the baseline. Figure \ref{fig:mars_hard_rel_hist} demonstrates visually that our method reduces the $\Delta V$ required for aerocapture on most of the hardest 2500 Monte Carlo cases, and on nearly all of the hardest 500 Monte Carlo cases.

\subsection{Aerocapture at Uranus}
Our Uranus aerocapture experiment uses the same initial state, target conditions, and vehicle parameters as Matz et al. \cite{matz2024analysis}. These parameters are presented in Table \ref{tab:uranus_parameters}. Because Uranus has a much larger planetary radius than Mars, we rescale the Uranus aerocapture problem in order to improve numerical stability.
\begin{table}[htbp]
	\fontsize{10}{10}\selectfont
    \caption{Initial and target conditions and vehicle parameters for Uranus aerocapture.}
    \label{tab:uranus_parameters}
        \centering 
   \begin{tabular}{|c|c|} 
      \hline 
      \textbf{Parameter} & \textbf{Value}\\
      \hline 
      Initial planet-relative altitude, $h_0$ (km) & 1000 \\
      Initial planet-relative velocity, $v_0$ (km/s) & 26.4\\
      Initial flight-path angle, $\gamma_0$ (deg) & -11.1\\
      Ballistic coefficient, $B_c$ (kg/m$^2$) & 180 \\
      Nominal L/D ratio & 0.25 \\
      Target apoapsis, $r_a^*$ (km) & 575559 \\
      Target periapsis, $r_p^*$ (km) & 29559\\
      Maximum cosine bank angle, $u_{\max}$ & $\cos(15^\circ)$ \\
      Minimum cosine bank angle, $u_{\min}$ & $\cos(165^\circ)$ \\
      \hline
   \end{tabular}
\end{table}

In order to make the Uranus aerocapture scenario more realistic, we use Fully Numerical Predictor-corrector Aerocapture Guidance (FNPAG)\cite{lu2015optimal} to get an initial control guess, with the maximum cosine bank angle set to $\cos(30^\circ)$ and the minimum cosine bank angle set to $\cos(150)$, and with $t_f = 750$ sec. Then, we refine the initial control guess using nonlinear local optimization with time discretization of $[0, 50, 100, 150, 200, \ldots, 500, 550, 600, 650, 700, 750]$ seconds. Because we initialize the control guess with FNPAG rather than zero control, Algorithm \ref{alg: successive_convexification} only requires 10 iterations to converge. 

\begin{figure}[t]
	\centering\includegraphics[width=2.5in,trim=0 0 0 20,clip]{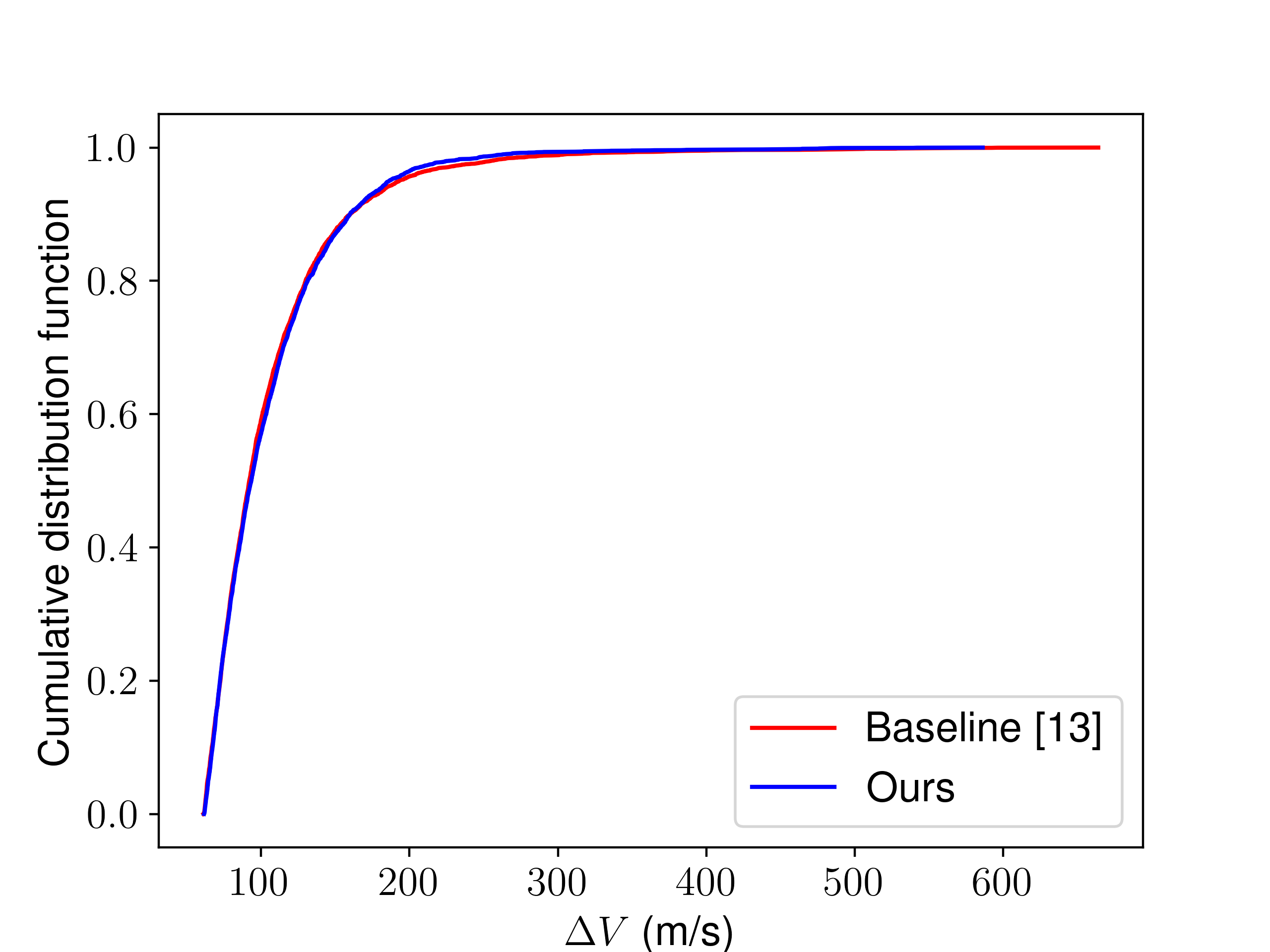}
    \includegraphics[width=2.5in,trim=0 0 0 20,clip]{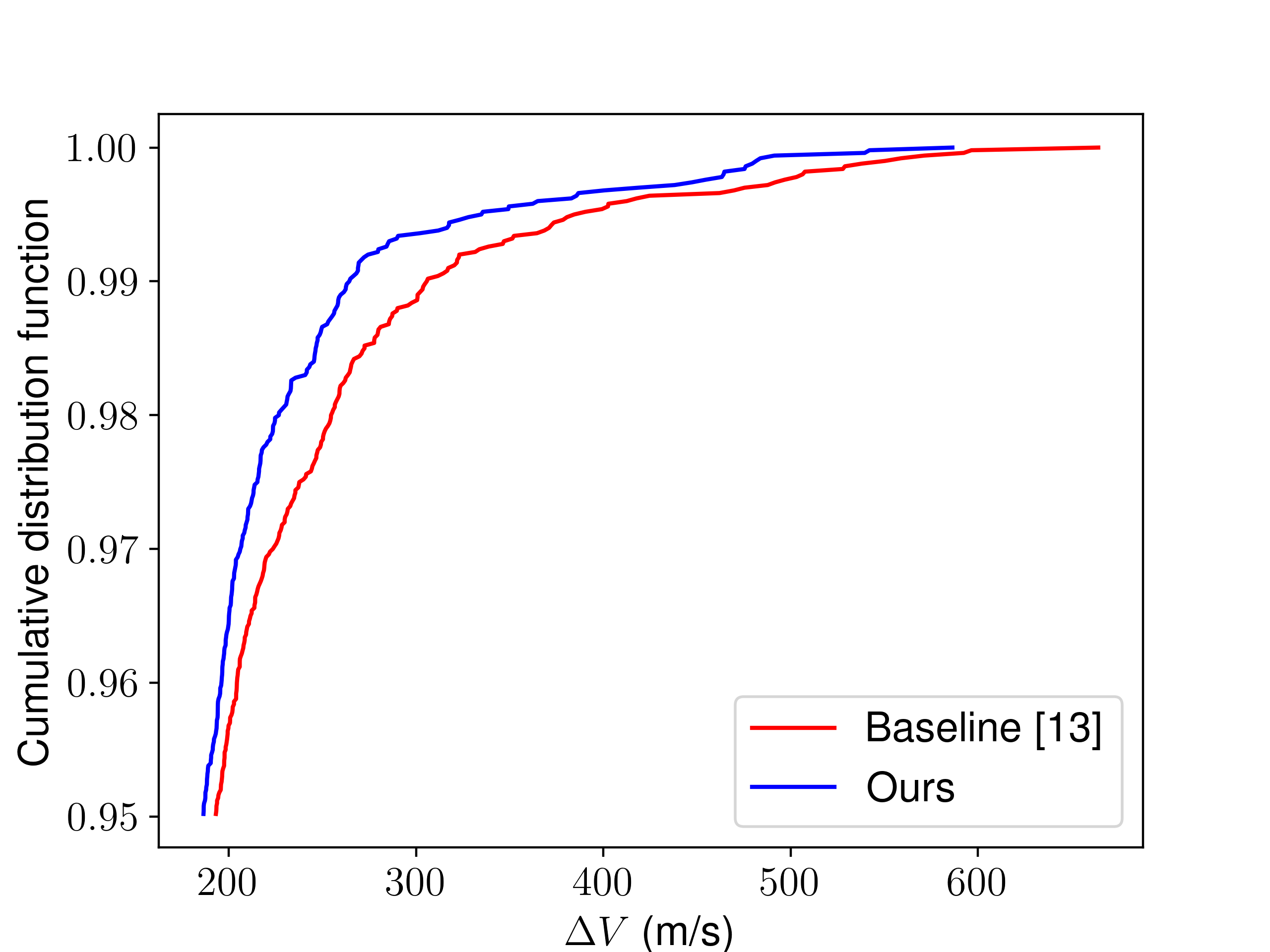}
	\caption{Full $\Delta V$ cumulative distribution function for our method and the baseline \cite{ridderhof2022chance} for Uranus aerocapture (left) and zoomed-in $\Delta V$ cumulative distribution function for Uranus aerocapture (right).}
	\label{fig:uranus_cdf}
%
	\centering\includegraphics[width=2.5in,trim=0 0 0 20,clip]{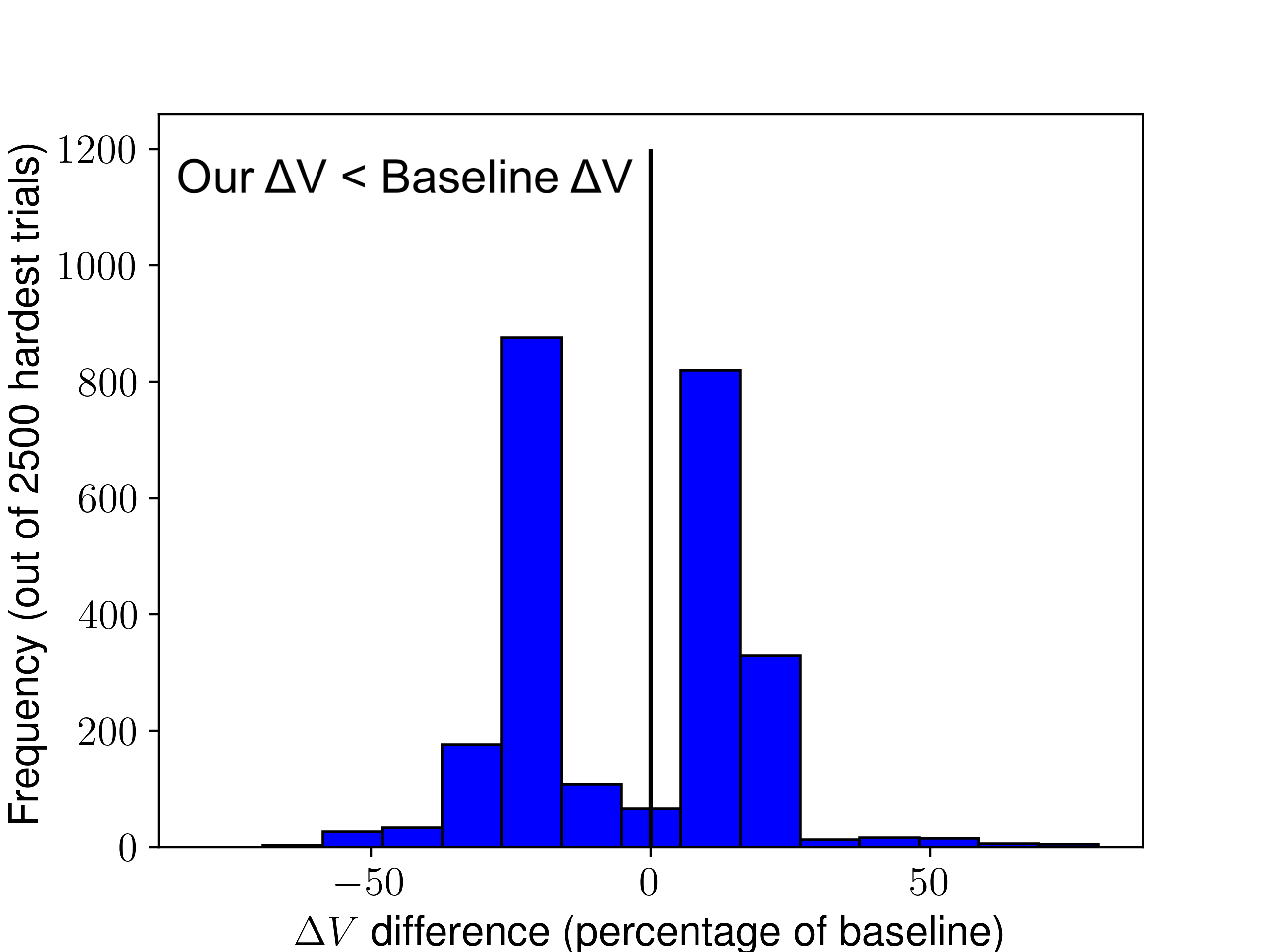}
    \includegraphics[width=2.55in,trim=0 0 0 20,clip]{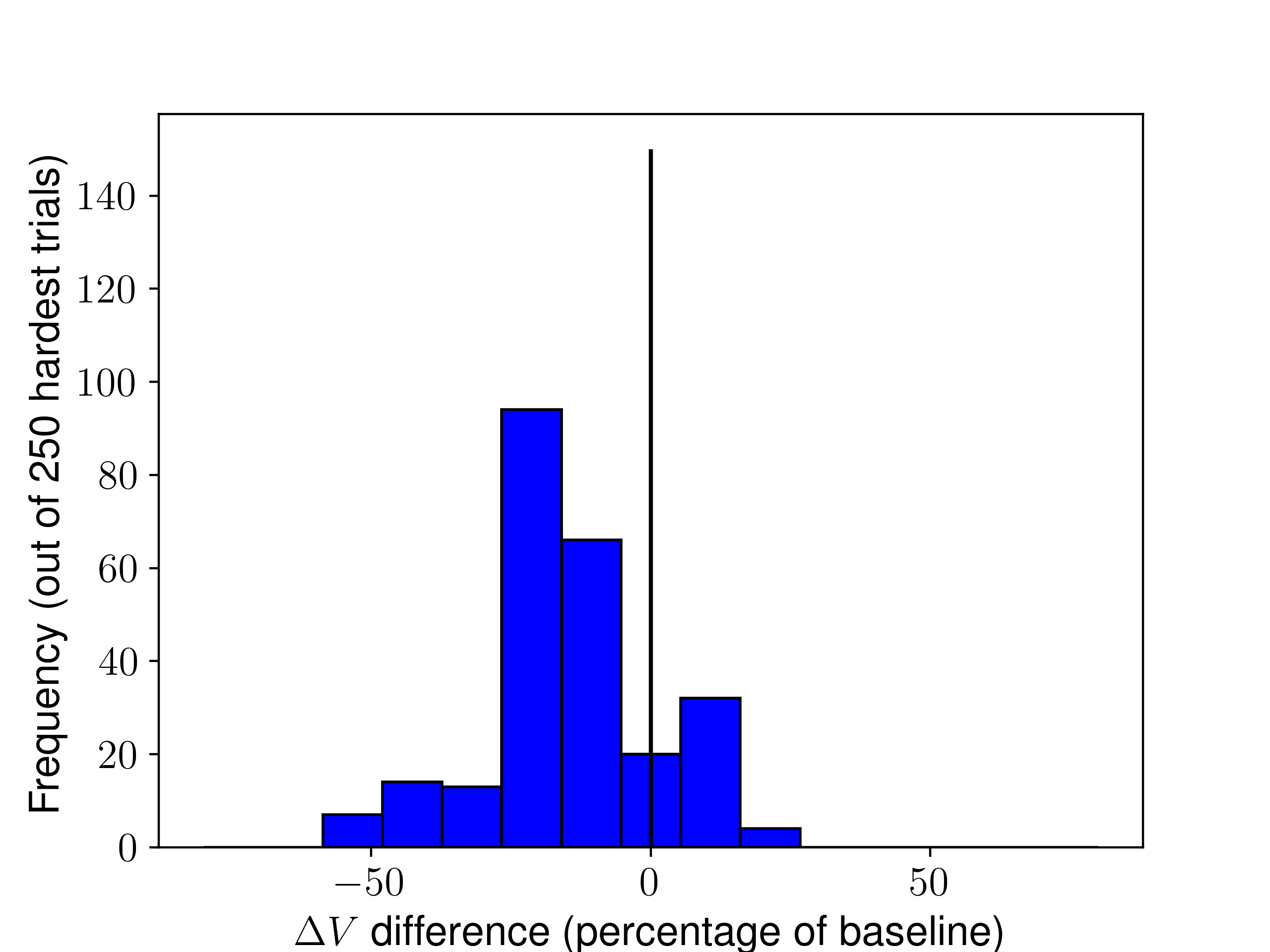}
	\caption{Difference between final $\Delta V$ for our method and the baseline \cite{ridderhof2022chance} for the hardest 2500 Uranus aerocapture cases (left) and the hardest 250 Uranus aerocapture cases (right).}
	\label{fig:uranus_rel_hist}
\end{figure}

We use UranusGRAM \cite{justh2024uranus} as our atmospheric model. We sample 1000 atmospheres from UranusGRAM and use the unbiased sample mean and covariance as the atmospheric density mean and covariance for both covariance steering algorithms, then sample an additional 5000 atmospheres from UranusGRAM for our Monte Carlo simulations. We also minimize 99.7th-percentile $\Delta V$ rather than 99th-percentile $\Delta V$ to better match NASA's safety requirements for aerocapture. We use a medium-difficulty initial velocity dispersion, with the $3\sigma$ initial state dispersion equal to 0 km in altitude, 0.2 km/s in the magnitude of the velocity, and 0.1 degrees in the flight path angle of the velocity vector. Matz et al. find that with our target orbit and entry conditions, making the entry flight path angle shallower by as little as $0.025^\circ$ leads to an increase in scenario difficulty, with an increase in the number of hyperbolic cases when using FNPAG for aerocapture guidance \cite{matz2024analysis}.
The Monte Carlo results are presented in Figures \ref{fig:uranus_cdf} and \ref{fig:uranus_rel_hist} and in Table \ref{tab:uranus_results}.

We find that our robust sampling-based covariance steering algorithm reduces worst-case $\Delta V$ and 99.7th percentile $\Delta V$ (our objective) by about 12\%, and reduces 99th percentile $\Delta V$ by 13.5\%. Our method also achieves a slight reduction in mean $\Delta V$. We find that our method has similar $\Delta V$ requirements to the baseline for the easiest 95\% of cases, but achieves meaningful $\Delta V$ reduction for the hardest 5\% of cases. Figure \ref{fig:uranus_rel_hist} demonstrates the strong performance of our robust sampling-based covariance steering algorithm on the hardest 250 Monte Carlo cases (5\% of the total 5000 cases). 

Our method generally yields slightly higher exit velocities than the baseline. As a result, the difference in $\Delta V$ between our method and the baseline is bimodal for the hardest 50\% of cases (as seen in Figure \ref{fig:uranus_rel_hist}). Our method tends to outperform the baseline in cases where the atmospheric and entry conditions slow down the aerocapture vehicle more than expected, while the baseline performs better in cases where the aerocapture vehicle slows down less than expected. Over the easiest 95\% of cases, these performance differences cancel out, and our method performs statistically similarly to the baseline, as seen in Figure \ref{fig:uranus_cdf} and Table \ref{tab:uranus_results}. In the hardest 5\% of cases, our method consistently outperforms the baseline, as seen in Figures \ref{fig:uranus_cdf} and \ref{fig:uranus_rel_hist}.

\begin{table}[hb]
	\fontsize{10}{10}\selectfont
    \caption{Median, 99th percentile, 99.7th percentile, and maximum $\Delta$V for Uranus aerocapture.}
    \label{tab:uranus_results}
        \centering 
   \begin{tabular}{|c|c|c|c|c|c|} 
      \hline 
      \textbf{Method}    & \textbf{Median} $\Delta$V & \textbf{Mean} $\Delta V$ & \textbf{99th pct.} $\Delta$V & \textbf{99.7th pct.} $\Delta$V & \textbf{Maximum} $\Delta$V\\
      \hline 
      Baseline \cite{ridderhof2022chance} & \textbf{92.0} & 106.7 & 305.7 & 475.4 & 664.3 \\
      Ours & 93.3 & \textbf{106.2} & \textbf{264.3} & \textbf{418.7} & \textbf{586.4} \\
      \hline
   \end{tabular}
\end{table}

\section{Conclusion}
In this work, we develop a new robust sampling-based covariance steering algorithm designed for aerocapture guidance. Our algorithm samples points from the entry state distribution, propagates nonlinear trajectories to atmospheric exit, and minimizes the worst-case 99th-percentile or 99.7th-percentile $\Delta V$ required for successful aerocapture over the collection of sampled trajectories. We apply our method to two Mars aerocapture scenarios with low atmospheric uncertainty and near-circular target orbits and one Uranus aerocapture scenario with high atmospheric uncertainty and an elliptical target orbit. We evaluate its performance relative to a state-of-the-art covariance steering method via Monte Carlo analysis, demonstrating a 5-15\% improvement in 99th-percentile, 99.7th percentile, and worst-case $\Delta V$ across scenarios without an increase in mean $\Delta V$.

\section{Future Work}
Updating the estimated atmosphere and estimated atmospheric uncertainty online and re-running our algorithm with the updated uncertainty estimate could improve performance, especially for scenarios where the atmosphere is initially highly uncertain. Our method could also be extended in the future to apply to scenarios with non-Gaussian initial state dispersions.

\section{Acknowledgment}
This work was supported by the National Science
Foundation Graduate Research Fellowship under grant no. 2141064. This work was also supported by the Draper Scholars program.


\bibliographystyle{AAS_publication}   
\bibliography{references}   

\end{document}